\documentclass[%
 reprint,
 amsmath,amssymb,
 aps,
pra,
superscriptaddress
]{revtex4-1}

\usepackage{graphicx}
\usepackage{bm}
\usepackage{braket}
\usepackage{array}
\usepackage{float}
\usepackage{epstopdf}
\usepackage{amssymb}	
\usepackage{mathrsfs}
\begin{document}

\title{Nondegenerate Two-Photon Absorption in GaAs/AlGaAs Multiple Quantum Well Waveguides}
\author{Nicholas Cox}
\affiliation{CREOL, The College of Optics and Photonics, University of Central Florida, Orlando, FL, 32816, USA}
\author{Junxiong Wei}
\affiliation{OPERA-Photonique, Universit\'e libre de Bruxelles (ULB), 50 Avenue F. D. Roosevelt, CP 194/5, 1050 Brussels, Belgium}
\author{Himansu Pattanaik}
\thanks{Present address: FARO Technologies, Londonderry, NH 03053, USA}
\affiliation{CREOL, The College of Optics and Photonics, University of Central Florida, Orlando, FL, 32816, USA}
\altaffiliation[text]{text}
\author{Thamer Tabbakh}
\affiliation{CREOL, The College of Optics and Photonics, University of Central Florida, Orlando, FL, 32816, USA}
\affiliation{King Abdulaziz City for Science and Technology, Riyadh 12354, Saudi Arabia}
\author{Simon-Pierre Gorza}
\affiliation{OPERA-Photonique, Universit\'e libre de Bruxelles (ULB), 50 Avenue F. D. Roosevelt, CP 194/5, 1050 Brussels, Belgium}
\author{David Hagan}
\email{hagan@creol.ucf.edu}
\author{Eric W. Van Stryland}
\affiliation{CREOL, The College of Optics and Photonics, University of Central Florida, Orlando, FL, 32816, USA}

\date{\today}

\begin{abstract}
    We present femtosecond pump-probe measurements of the nondegenerate ($1960$\,nm excitation and $1176$--$1326$\,nm probe) two-photon absorption spectra of $8 \,\mathrm{nm}$ GaAs/$12$\,nm Al$_{0.32}$Ga$_{0.68}$As quantum well waveguides. Experiments were performed with light pulses co-polarized normal and tangential to the quantum well plane. The results are compared to perturbative calculations of transition rates between states determined by the $\mathbf{k}\cdot\mathbf{p}$ method with an 8 or 14 band basis. We find excellent agreement between theory and experiment for normal polarization, then use the model to support predictions of orders-of-magnitude enhancement of nondegenerate two-photon absorption as one constituent photon energy nears an intersubband resonance.
\end{abstract}

\pacs{Valid PACS appear here}
\maketitle

\section{Introduction} Nondegenerate two-photon absorption (ND-2PA) is a process whereby absorption of an optical field is induced by a second, high irradiance field at a different wavelength. Applications of ND-2PA include detection \cite{Fishman2011}, imaging \cite{Pattanaik2016a}, and all-optical switching \cite{Liang2005}. Inverting carrier populations can also transform ND-2PA into nondegenerate two-photon gain \cite{Reichert2016a, Melzer2018a, Hayat2008}, which is critical for realizing a two-photon semiconductor laser \cite{Ironside1992, Gauthier1992, Hayat2011}. Waveguides are especially interesting for nonlinear optical applications because they enable strong effects through long interaction lengths. Group velocity mismatch (GVM) induced walkoff usually limits nondegenerate interactions, but dispersion engineering can mitigate or remove this walkoff entirely \cite{Poulvellarie2018}.

The degenerate 2PA (D-2PA) spectrum of infinite quantum wells was first predicted by \citet{Spector1987b} and \citet{Pasquarello1988}. Shortly after, \citet{Nithisoontorn1989} experimentally demonstrated D-2PA to excitons in GaAs quantum wells. \citet{Shimizu1989} developed an excitonic model for D-2PA, and \citet{Tai1989} verified their predictions with two-photon luminescence spectra. Later, \citet{Yang1993} showed the anisotropy of D-2PA in quantum well waveguides.

\citet{Pasquarello1990a} relaxed some of Shimizu's approximations and extended the analysis to nondegenerate photon pairs, predicting large enhancements as one photon energy neared an intersubband resonance. \citet{Pattanaik2016} quantitatively examined these nondegenerate resonance enhancements, expanding upon a six-band theory developed by \citet{Khurgin1994}.

Here, we present the first pump-probe measurements of ND-2PA coefficients in GaAs quantum wells. We studied $8$\,nm GaAs/$12$\,nm Al$_{0.32}$Ga$_{0.68}$As quantum wells at room temperature using beams polarized normal (TM-TM) and tangential (TE-TE) to the quantum well plane, and compared the results with a theoretical model for ND-2PA in finite wells neglecting excitonic effects. 

We find that our perturbative model matches experimental results very closely for TM-TM beams, whereas a relatively large error in TE-TE predictions indicates that a more thorough analysis is needed. The TM-TM model shows that predictions of intersubband resonance enhancements of ND-2PA also apply to finite wells, suggesting the possibility of extremely sensitive gated detection of sub-bandgap pulses \cite{Fishman2011}.

This article is organized as follows. In Sec.\,\ref{sec:Background}, we derive a method for calculating ND-2PA coefficients in an arbitrary quasi-2D semiconductor. We introduce our GaAs quantum well waveguide in Sec.\,\ref{sec:Experiment} and describe the pump-probe experiments carried out to find its ND-2PA coefficients. In Sec.\,\ref{sec:Calculation}, we apply the model of Sec.\,\ref{sec:Background} to a simple quantum well structure that approximates our sample. Measurement results are presented in Sec.\,\ref{sec:Results} followed by a discussion in Sec.\,\ref{sec:Discussion}. 
    
Appendix \ref{sec:BulkBandStructure} contains background information about the Kane band structure model for zinc blende semiconductors. Appendix \ref{sec:IntersubbandElement} shows the derivation of an intersubband matrix element in the envelope function expansion. In Appendix \ref{sec:Material}, we give the equations and parameters used in numerical simulations of quantum well states and optical modes. Appendix \ref{sec:Propagation} contains nonlinear wave propagation analysis, as well as the techniques used to match the calculation results to experimental curves. Less essential equation derivations are placed in the supplemental material \cite{Supp}.

\section{Theoretical Background}\label{sec:Background}

The $n$-th level wave function of the $j$-th band (e.g. conduction, heavy hole, light hole) in a semiconductor confined in the $z$ direction is \cite{Chuang1995}
\begin{equation}\label{eq:state}
    \psi_{jn}(\mathbf{r};\mathbf{k}_t) = e^{i\mathbf{k}_t\cdot\mathbf{r}_t}F_{jn}(\mathbf{r}_t, z; \mathbf{k}_t).
\end{equation}
$F_{jn}$ is an envelope with lattice periodicity only in $\mathbf{r}_t$, the component tangential to the quantum well plane.

Second order perturbation theory gives the net two-photon transition rate per unit volume \cite{Lee1974}
\begin{align}\label{eq:WcvRate}
    W &= \frac{2\pi}{\hbar}\frac{1}{V}\sum_{cv}\sum_{\mathbf{k}_t} \nonumber \\
    &\left|\sum_i \frac{\braket{c|\hat{H}_2'|i}\braket{i|\hat{H}_1'|v}}{E_{iv}(\mathbf{k}_t) - \hbar\omega_1} + \frac{\braket{c|\hat{H}_1'|i}\braket{i|\hat{H}_2'|v}}{E_{iv}(\mathbf{k}_t) - \hbar\omega_2} \right|^2 \nonumber \\
    &\times\delta[E_{cv}(\mathbf{k}_t) - \hbar\omega_1 -\hbar\omega_2],
\end{align}
where $\ket{c}$ and $\ket{v}$ are conduction and valence envelopes, respectively. $H'_l$ is the interaction Hamiltonian for a vector potential of magnitude $A_{0l}$ and polarization $\hat{\mathbf{e}}_l$, given by \cite{Basov1966}
\begin{equation}\label{eq:EMHamiltonian}
    \hat{H}_l' = \frac{e A_{0l}}{2 m_0}\hat{\mathbf{e}}_l\cdot \left(\hat{\mathbf{p}} + \hbar\mathbf{k}_t \right).
\end{equation}
The $\hbar\mathbf{k}_t$ term arises from the chain rule for the momentum operator $\hat{\mathbf{p}} = -i\hbar\nabla$ applied to states in the form of Eq.\,(\ref{eq:state}). This term is usually ignored because it frequently cancels out, but we leave it in for completeness.

The transition rate is converted to an ND-2PA coefficient by \cite{Sheik-Bahae1998}
\begin{equation}\label{eq:2PACoef}
    \alpha_2(\omega_1;\omega_2) = \frac{\hbar\omega_1}{2I_1 I_2}W,
\end{equation}
which describes the attenuation of wave $1$ at frequency $\omega_1$ induced by wave $2$ at $\omega_2$. $I_1$ and $I_2$ are the incident field irradiances
\begin{equation}\label{eq:Irradiance}
    I_l = \frac{1}{2}n_lc\epsilon_0 \omega_l^2|A_{0l}|^2,
\end{equation}
with $n_l$ the effective index at $\omega_l$. We also introduce a unitless matrix element between envelopes \cite{Hutchings1992c}
\begin{equation}\label{eq:normalizedM}
    M_{jn,im}^{(l)}(\mathbf{k}_t) = \frac{\hbar}{m_0 P} \hat{\mathbf{e}}_l\cdot\braket{F_{jn}|\mathbf{p} + \hbar\mathbf{k}_t |F_{im}}.
\end{equation}
As detailed in Appendix \ref{sec:BulkBandStructure}, the Kane parameter $P = \hbar/m_0\braket{iS|p_x|X}$ is the optical coupling strength between conduction and valence bands. 

Finally, we combine Eqs.\,(\ref{eq:WcvRate})--(\ref{eq:normalizedM}) into a general expression for ND-2PA coefficients: \cite{Supp} 
\begin{equation}\label{eq:2PAGeneral}
    \alpha_2(\omega_1;\omega_2) = K\frac{E_p}{n_1n_2L_zE_g^4}f_2\left(\frac{\hbar\omega_1}{E_g}; \frac{\hbar\omega_2}{E_g}\right),
\end{equation}
where $f_2$ is the dimensionless spectral function
\begin{align}\label{eq:SpectralFunction}
    f_2(x_1;x_2) &= \sum_{cv}\sum_{\kappa_0}\frac{1}{2\pi}\int_0^{2\pi}\kappa_0\left|\frac{\partial{\epsilon_{cv}}}{\partial{\kappa}}\right|^{-1}_{\kappa_0} \nonumber \\ 
                 &\times\left|\sum_{i}\frac{M_{ci}^{(2)}M_{iv}^{(1)}}{\epsilon_{iv} - x_1} + \frac{M_{ci}^{(1)}M_{iv}^{(2)}}{\epsilon_{iv} - x_2}\right|d\phi.
\end{align}

The quantity $E_g$ is the bandgap of the quantum well material and $E_p = 2m_0 P^2/\hbar^2$ is the Kane energy. The parameter $L_z$ is the total thickness of the structure in the $z$ direction. For a single quantum well, $L_z$ is the sum of the barrier and well widths.

Eq.\,(\ref{eq:2PAGeneral}) is valid in any unit system so long as the material-independent parameter $K$ is adjusted accordingly. With energies and lengths written in Hartree atomic units ($\hbar = m_0 = e = 1/(4\pi\epsilon_0) = 1$), this constant is simply $K = (\pi/c)^2 = (\pi/137)^2$. The final 2PA coefficient can then be converted to cm/GW by the conversion factor $1 \,\mathrm{au} = 29.36 \,\mathrm{cm/GW}$.

The integral in Eq.\,(\ref{eq:SpectralFunction}) is taken over the azimuthal angle of $\mathbf{k}_t$, whose magnitude has been replaced by the unitless quantity $\kappa = k_t P/E_g$. Energies are also normalized by letting $\epsilon_{jk} = E_{jk}/E_g$ and $x_j = \hbar\omega_j/E_g$. Each $\kappa_0$ is a real, positive solution to
\begin{equation}\label{eq:resonance}
    \epsilon_{cv}(\kappa, \phi) - x_1 - x_2 = 0.
\end{equation}

We derived Eqs.\,(\ref{eq:2PAGeneral}) and (\ref{eq:SpectralFunction}) without making any assumptions of specific band structure or layer design, leaving us with a general expression for ND-2PA coefficients in quasi-2D materials. Later, we make approximations to simplify calculations for the symmetric quantum wells introduced in the next section.

\section{Sample and Experiment}\label{sec:Experiment}
\begin{figure}[htb]
    \centering
    \includegraphics{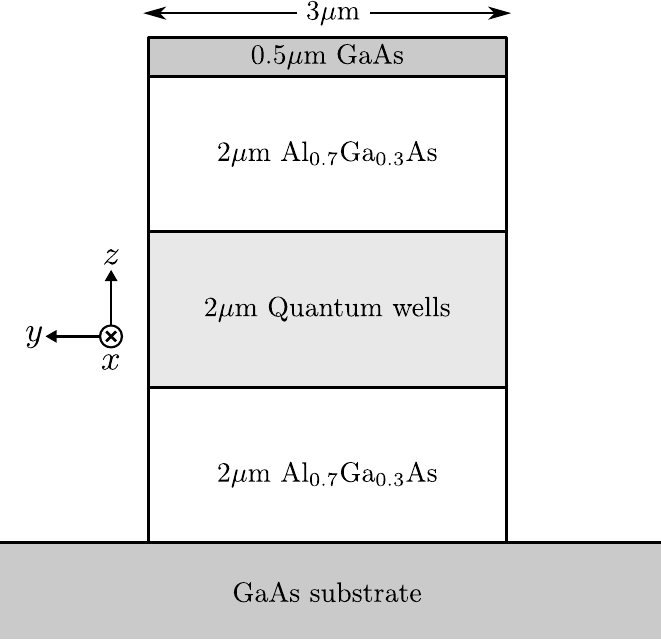}
    \caption{A schematic of the quantum well waveguide samples studied. The $2 \,\mathrm{\mu m}$ quantum well region consists of $100\times$ ($8 \,\mathrm{nm}$ GaAs)/($12 \,\mathrm{nm}$ $\mathrm{Al_{0.32}Ga_{0.68}}$) quantum wells.}
    \label{fig:Sample}
\end{figure}

We experimentally investigated GaAs quantum wells MBE-grown by Sandia National Laboratories on an n-GaAs (100) substrate. A wave guiding structure was formed in the $z$ direction by growing $2\,\mu\mathrm{m}$ thick Al$_{0.7}$Ga$_{0.3}$As cladding layers on either side of a $2\,\mathrm{\mu m}$ active region. The active region comprised $100$ repetitions of ($8 \,\mathrm{nm}$ GaAs)/($12 \,\mathrm{nm}$ Al$_{0.32}$Ga$_{0.68}$As) quantum wells, with barrier widths chosen so that coupling between wells is negligible. Transverse optical confinement was achieved by etching a $3\,\mathrm{\mu m}$ wide ridge through the active region and lower cladding. Finally, the sample was cleaved to a length of $3.6 \,\mathrm{mm}$. The layer structure and geometry are seen in Fig.\,\ref{fig:Sample}. 

\begin{figure}[htb]
     \centering
     \includegraphics{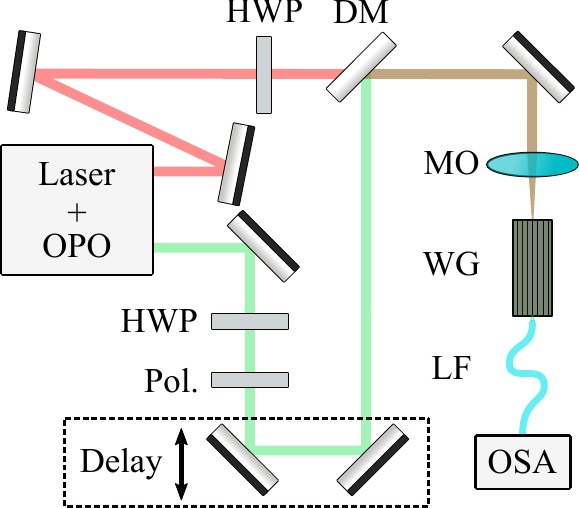}
     \caption{A schematic of the experimental setup. OPO: optical parametric oscillator, HWP: half-wave plate, Pol: Polarizer, DM: dichroic mirror, MO: microscope objective, WG: AlGaAs/GaAs quantum well waveguide, LF: lensed fiber, OSA: optical spectrum analyzer. The pump laser for the OPO is a Ti:Sapphire laser which is tuned between $730 \,\mathrm{nm}$ and $795 \,\mathrm{nm}$}
     \label{fig:ExperimentalSetup}
 \end{figure}

Fig.\,\ref{fig:ExperimentalSetup} shows the optical setup employed to study the sample. A short wavelength probe and long wavelength pump came from the signal and idler, respectively, of a Spectra-Physics OPAL optical parametric oscillator (OPO) synchronously driven by a Spectra-Physics Tsunami Ti:Al$_2$O$_3$ laser with $82$MHz repetition rate. We tuned the driving laser wavelength between $730 \,\mathrm{nm}$ and $795 \,\mathrm{nm}$ to study 2PA at sum photon energies near the absorption edge. For each driving laser wavelength, the OPO phase matching was adjusted to fix the idler at $1960 \,\mathrm{nm}$. In effect, the pump was fixed at $1960 \,\mathrm{nm}$ while the probe varied between $1176 \,\mathrm{nm}$ and $1326 \,\mathrm{nm}$. This pump photon energy was chosen to be below the D-2PA edge so that two-photon photogenerated carriers did not interfere with data interpretation. 

 After fixing the signal and idler wavelengths, their polarizations were set to TE ($y$-polarized) or TM ($z$-polarized) using broadband half-wave plates. The probe then traveled through a delay line and combined with the pump at a dichroic mirror. The beams were end-fire coupled into the ridge waveguide (Fig.\,\ref{fig:Sample}) using a microscope objective and collected by a lensed fiber at the exit facet. The lensed fiber was connected to a Yokogawa AQ6370D spectrum analyzer (OSA in Fig.\,\ref{fig:ExperimentalSetup}) to compare the probe output power spectrum with and without the pump's presence. The process was repeated at a series of probe delays to generate curves of normalized transmission versus delay.

 It is necessary to know the pump power at the facet inside the waveguide to convert normalized transmission to an ND-2PA coefficient. This input power was calculated by back-propagating the OSA-measured output power to the front facet using 
 \begin{equation}
     P_{out} = P_{in}\exp(-\sigma L)T,
 \end{equation}
which depends on the waveguide loss $\sigma$, the sample propagation length $L$, and the the output facet coupling efficiency $T$. This efficiency was experimentally determined by temporarily replacing the microscope objective of Fig.\,\ref{fig:ExperimentalSetup} with a lensed fiber identical to the one at the output. By measuring $P_{out}/P_{in}$ for this symmetric system, we found the facet transmission by 
 \begin{equation}
     T = \left[\frac{P_{out}}{P_{in} }\exp(\sigma L)\right]^{1/2}.
 \end{equation}
 After finding a waveguide loss of $\sigma_{TM} = 0.46\,\mathrm{mm}^{-1}$ and $\sigma_{TE} = 0.56\,\mathrm{mm}^{-1}$ (See Appendix \ref{subsec:fitting}), we determined the transmission coefficients of the front facet to be $T_{TM} = 0.47$ and $T_{TE} = 0.54$.

Autocorrelation measurements of the $1960 \,\mathrm{nm}$ pump at four different sum wavelengths gave the following pulsewidths: $227 \,\mathrm{fs}$ at $789 \,\mathrm{nm}$, $221 \,\mathrm{fs}$ at $774 \,\mathrm{nm}$, $156 \,\mathrm{fs}$ at $754 \,\mathrm{nm}$, and $149 \,\mathrm{fs}$ at $745 \,\mathrm{nm}$. By comparing to the measured spectra, we determined pump pulsewidths to be an average of $9\%$ greater than the Gaussian bandwidth limit.
\section{Calculation of 2PA coefficients}\label{sec:Calculation}
This section describes the model used to calculate the 2PA coefficients of a symmetric GaAs quantum well. We begin by calculating the energy levels and envelope functions of each subband level. Then we construct expressions for the optical matrix elements between all states. Finally, Eq.\,(\ref{eq:SpectralFunction}) is used to calculate the 2PA coefficients for parabolic bands.
\subsection{Wavefunction envelopes}\label{ssec:Envelopes}
Calculation of ND-2PA coefficients requires knowledge of the energy levels and wave functions between which two-photon transitions occur. We begin by expanding the envelope functions in the basis of zone center wavefunctions $u_{\nu 0}(\mathbf{r})$: \cite{Kohn1954, Bastard1981}
\begin{equation}\label{eq:envExp}
    F_{jn}(\mathbf{r}, \mathbf{k}_t) = \sum_{\nu}\chi_{jn}^{\nu}(z; \mathbf{k}_t)u_{\nu 0}(\mathbf{r}).
\end{equation}
The basis is chosen to consist of either 8 or 14 spin-degenerate bands (see Appendix \ref{sec:BulkBandStructure} for details). For ease of calculation, we only solve for envelopes at $\mathbf{k}_t = 0$; the approximation applied for $\mathbf{k}_t \neq 0$ is discussed later in this subsection. Taking the alloy composition-dependent energy offset of band $j$ as a z-dependent potential $V_j(z)$ leads to a second order Schr{\"o}dinger Equation \cite{Bastard1976}

\begin{equation}\label{eq:SEM}
    \frac{1}{2}p_z \frac{1}{m_{j}^z(E_{jn},z)}p_z\chi_{jn}^j(z) + V_{j}(z)\chi_{jn}^j(z) = E_{jn}\chi_{jn}^j(z).
\end{equation}
The superscript on $\chi_{jn}^j$ denotes the dominant envelope, and $m_j^z(E_{jn},z)$ is the state's energy-dependent effective mass in the $z$ direction. Choosing the 8 band basis for Eq.\,(\ref{eq:envExp}) yields the effective mass relation \cite{Bastard1976, Supp}

\begin{align}\label{eq:mc}
    \frac{m_0}{m_c^z(E,z)} &= 1 + \frac{m_0}{\mathscr{M}_{cc}^{zz}} + \frac{2}{3}\frac{E_p}{E - V_{l}(z)} \nonumber \\
                           &+ \frac{1}{3}\frac{E_p}{E + \Delta - V_s(z)} \nonumber \\
\frac{m_0}{m^z_{l}(E,z)} &= 1 + \frac{m_0}{\mathscr{M}_{ll}^{zz}} + \frac{2}{3}\frac{E_p}{E - V_{c}(z) - E_g} \nonumber \\
\frac{m_0}{m^z_{h}(E,z)} &= 1 + \frac{m_0}{\mathscr{M}_{hh}^{zz}}.
\end{align}
The quantity $\Delta$ is the spin-orbit split-off energy and $m_0/\mathscr{M}_{kk}^{zz}$ are remote band contributions included by L{\"o}wdin's perturbation method \cite{Lowdin1951}. We take the approximation that this remote band contribution is independent of energy.

In the 14 band basis, the conduction band effective mass changes to \cite{Supp}
\begin{align}\label{eq:mc2}
    \frac{m_0}{m_c^z(E,z)} &= 1 + \frac{m_0}{\mathscr{M}_{cc}^{zz}} \nonumber \\
    &+\frac{2}{3}\frac{E_p}{E - V_{l}(z)} + \frac{1}{3}\frac{E_p}{E + \Delta - V_s(z)} \nonumber \\
    &+ \frac{2}{3}\frac{E_p'}{E - E_{g}' - V_{l'}(z)} + \frac{1}{3}\frac{E_p'}{E - E_{s'} - V_{s'}(z)},
\end{align}
where $l'$ is the light electron band at $E = E_g'$ and $s'$ is the split-off electron band at $E = E_{s'}$. $E_p'$ is the coupling energy between the two sets of conduction bands. All hole effective masses are identical to the 8 band results.

If we set $V_j = 0$ and $E = E_{j,\mathrm{bulk}}$ in Eqs.\,(\ref{eq:mc}) and (\ref{eq:mc2}), we find expressions for bulk GaAs effective masses. As was done in Ref.\,\cite{Meney1994}, we choose $m_0/\mathscr{M}_{kk}^{zz}$ so that calculated bulk effective masses match experimental values.

We assume the $\mathbf{k}_t$ band dispersion to have the form
\begin{equation}\label{eq:papx}
    E_{jn}(\mathbf{k}_t) = E_{jn}(0) + \frac{\hbar^2 k_t^2}{2m^t_j},
\end{equation}
where $m_j^t$ is the effective mass in the transverse direction. This parabolic band approximation has been successfully used to calculate 2PA coefficients for bulk semiconductors \cite{Basov1966}, at the cost of ignoring some fine structure in the dispersion \cite{Hutchings1992c}. 

\subsection{Matrix Elements}
Equipped with a model for electronic states, we can proceed to calculate the optical matrix elements between them. Two-photon transitions across the bandgap always require an interband transition, for which we consider allowed and forbidden paths. For TM-TM polarizations, the remaining step is assumed to be an allowed intersubband transition. In TE-TE 2PA, the remaining step is assumed to be a forbidden self transition \cite{Wherrett1984a}. These two interaction types vary differently with wavelength near the 2PA edge, causing anisotropy between the polarization schemes.

Intersubband matrix elements are calculated between the envelopes found from Eq.\,(\ref{eq:SEM}). As a consequence, the results are only strictly valid at the band edge. For TM polarization, we find that 
\begin{equation}\label{eq:Misb}
    M^z_{jn,im} = \frac{\hbar}{2P}\Braket{\chi_{jn}^j|\frac{1}{m'_i}p_z + p_z \frac{1}{m'_j}|\chi_{im}^j}\,\delta_{ij},
\end{equation}
where
\begin{equation}
    \frac{1}{m_k'} = \frac{1}{m_k(E, z)} - \frac{1}{\mathscr{M}_{kk}^{zz}}.
\end{equation}
See Appendix \ref{sec:IntersubbandElement} for justification of the above equation. This form exhibits two improvements over that in Refs. \cite{Pattanaik2016} and \cite{Khurgin1994}, \textit{viz}. $M_{jn,im}^z = \hbar/(m_j P)\braket{\chi^j_{jn}|p_z|\chi^j_{jm}}$. The first is to include energy scaling of the effective mass, and the second is to account for the fact  that interband coupling depends only on the inverse effective mass component arising from interactions within the basis. Eq.\,(\ref{eq:Misb}) can also be compared to that in Ref.\,\cite{[{See chapter 4 of }] Faist2013}, which ignores the subtraction of remote band contributions.

Self transitions are forbidden because they describe optical coupling between states of nearly identical symmetry. Eq.\,(\ref{eq:Misb}) with $m = n$ shows that these contributions are negligible for TM polarization. For TE fields we use the relation $\braket{\mathbf{p}} = (m_0/\hbar)\nabla_\mathbf{k} E(\mathbf{k}_t)- \hbar\mathbf{k}_t$ \cite{NeilW.Ashcroft1976} with the energy given by Eq.\,(\ref{eq:papx}). Fixing TE polarization in the $y$ direction leads to
\begin{equation}
    M^y_{jn,im} = \frac{\hbar^2}{m^t_j P}k_t\sin\phi\,\delta_{mn}\delta_{ij}.
\end{equation}

Eq.\,(\ref{eq:Misb}) is also valid for interband transitions, but using it would ignore the $\mathbf{k}_t$ dependence once again. Instead, we use the method of \citet{Yamanishi1984} to estimate matrix elements from the bulk band structure: 
\begin{equation}\label{eq:interband}
    M^{(l)}_{jn,im} = \frac{\hbar}{m_0P}\hat{\mathbf{e}}_l\cdot\braket{u_{j0}(\mathbf{r})|\hat{\mathbf{p}}|u_{i0}(\mathbf{r})}'\delta_{nm},
\end{equation}
where $u_{j0}(\mathbf{r})$ is the zone-center basis function for band $j$. As described in Appendix \ref{sec:BulkBandStructure}, the prime denotes that the basis functions are rotated by an angle $\theta = \cos^{-1}{[k_z/(k_z^2 + k_t^2)^{1/2}]}$. We find $k_z$ from the energy relation $E_{jn}(0) = \hbar^2 k_z^2/2m_j^z$. Eq.\,(\ref{eq:interband}) applies to TM and TE polarizations and accounts for allowed ($\propto\cos\theta$) and forbidden ($\propto\sin\theta$) transitions. 

All matrix elements for both polarizations are compiled in Tables \ref{tab:ZMatrixElements} and \ref{tab:YMatrixElements}, in a similar form to those in Refs.\,\cite{Lee1974} and \cite{Hutchings1992c}. 
\subsection{2PA coefficients for Parabolic Bands}

The energy separation between parabolic conduction and valence bands is given in normalized form by
\begin{equation}\label{eq:EnergySeparation}
    \epsilon_{cv}(\kappa) = \epsilon_{cv}(0) + \frac{m_0}{\mu_{cv}^t}\frac{\kappa^2}{\epsilon_p},
\end{equation}
where $\epsilon_p = E_p/E_g$ and $1/\mu_{cv}^t = 1/m_c^t - 1/m_v^t$. We see immediately from Eq.\,(\ref{eq:EnergySeparation}) that
\begin{equation}\label{eq:ParabolicJDOS}
    \left|\frac{\partial \epsilon_{cv}}{\partial \kappa}\right|^{-1}_{\kappa_0} = \frac{\mu_{cv}^t}{m_0}\frac{\epsilon_p}{2\kappa_0}.
\end{equation}
Combining Eqs.\,(\ref{eq:resonance}) and (\ref{eq:EnergySeparation}), we find 
\begin{equation}\label{eq:ParabolicK0}
    \kappa_0 = \left\{\frac{\mu_{cv}^t}{m_0}\epsilon_p \left[x_1 + x_2 - \epsilon_{cv}(0)\right]\right\}^{1/2}.
\end{equation}
Feeding the results of Eqs.\,(\ref{eq:EnergySeparation})--(\ref{eq:ParabolicK0}) into Eq.\,(\ref{eq:SpectralFunction}) gives a general expression for the dimensionless scaling factor in the parabolic band approximation: 
\begin{align}\label{eq:fgen}
    f_2(x_1&; x_2) = \epsilon_p\sum_{c\alpha, v} \frac{\mu_{cv}^t}{m_0}\Theta[x_1 + x_2 - \epsilon_{cv}(0)] \nonumber \\
                  &\times\frac{1}{2\pi}\int_{0}^{2\pi}\left|\sum_{i}\frac{M_{ci}^{(2)}M_{iv}^{(1)}}{\epsilon_{iv} - x_1} + \frac{M_{ci}^{(1)}M_{iv}^{(2)}}{\epsilon_{iv} - x_2}\right|^2 d\phi.
\end{align}
The sum runs only over $\alpha$ conduction states, with identical $c\beta$ transitions included by a factor of $2$. The step function $\Theta$ defines the range where the solution of Eq.\,(\ref{eq:ParabolicK0}) is real. The aforementioned angular rotation factor in the interband matrix element is found to be
\begin{equation}\label{eq:AngularFactor}
\cos\theta = \left(1 + \frac{\mu_{cv}^t}{m_c^{z}}\frac{x_1 + x_2 - \epsilon_{cv}(0)}{\epsilon_c(0)} \right)^{-1/2}.
\end{equation}
Two-photon absorption coefficients can finally be found from Eq.\,(\ref{eq:2PAGeneral}). Note that each quantum well is treated as a separate system so that $L_z$ in this equation is the total thickness of a single well (20\,nm).

Eq.\,(\ref{eq:fgen}) is valid for TM-TM, TE-TE and the mixed-polarization TE-TM configurations. Because the co-polarized schemes provide sufficient information about the ND-2PA anisotropy, we do not perform the less tractable cross-polarized calculations. The next two subsections provide some simplifications for the TM-TM and TE-TE cases.
\subsubsection{TM-TM 2PA coefficients}
\begin{figure}
    \centering
    \includegraphics{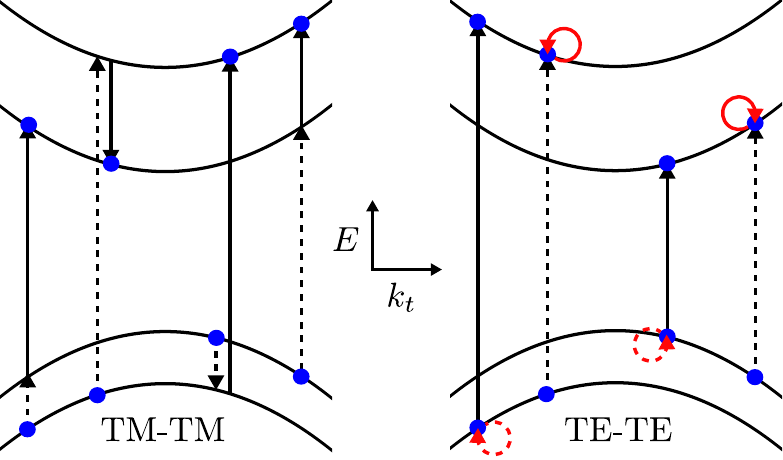}
    \caption{A diagram of the possible paths for TM-TM (left) and TE-TE (right) two-photon transitions. The initial and final states are marked with a blue dot. Dashed lines represent a non-resonant transition, which is the transition leading to a detuning denominator. Black and red lines signify allowed and forbidden transitions, respectively.}
    \label{fig:Transitions}
\end{figure}

As shown in Fig.\,\ref{fig:Transitions}, every TM-TM two-photon transition includes one interband and one intersubband transition. Because envelope parity alternates with subband index, the required TM matrix element (Eq.\,(\ref{eq:Misb})) imposes the selection rule $n - m = 2k + 1$ for integer $k$. Since every element in Table \ref{tab:ZMatrixElements} is independent of $\phi$, the azimuthal integral of Eq.\,(\ref{eq:fgen}) reduces to unity such that
\begin{align}\label{eq:TMf}
    f_2(x_1; x_2) &= \frac{\epsilon_p}{x_1 x_2^2}\sum_{c\alpha, v} \frac{\mu_{cv}^t}{m_0}\left|\sum_{i}\frac{M_{ci}^{z}M_{iv}^{z}}{\epsilon_{iv} - x_1} + \frac{M_{ci}^{z}M_{iv}^{z}}{\epsilon_{iv} - x_2}\right|^2 \nonumber \\
                  &\times\Theta[x_1 + x_2 - \epsilon_{cv}(0)].
\end{align}
Both matrix elements are non-zero at the ND-2PA edge for doubly-allowed transitions, giving the step-like shape characteristic of linear quantum well absorption.

\subsubsection{TE-TE 2PA coefficients}
Per Fig.\,\ref{fig:Transitions}, each TE-TE 2PA path involves an interband and a self transition. Two-photon transitions therefore inherit the selection rules of the the interband transition, namely $m = n$. Using the matrix elements from Table \ref{tab:YMatrixElements}, which are generally $\phi$-dependent, we simplify the dimensionless scaling function to 
\begin{align}\label{eq:TEalpha22}
    f_2(x_1; x_2) &= \frac{4}{x_1 x_2^2}\left(\frac{1}{x_1} + \frac{1}{x_2}\right)^2\sum_{c\alpha,v}\Braket{\left|M^y_{c\alpha,v}\sin\phi \right|^2}_{\phi}\nonumber \\
                  &\times\left[x_1 + x_2  - \epsilon_{cv}(0)\right] \Theta[x_1 + x_2 - \epsilon_{cv}(0)]. \nonumber \\
\end{align}
The term linear in $x_1 + x_2$ grows from zero at the 2PA edge, meaning TE-TE 2PA dispersion lacks the discontinuities seen in TM-TM 2PA curves. 
\begin{table*}[p]
    \caption{\label{tab:ZMatrixElements} Optical matrix elements $M^z_{jn,im} = \hat{\mathbf{z}}\cdot\mathbf{M}_{jn,im}$. Only transitions to $\alpha$ spin states are shown since transitions to $\beta$ states can be found from the relation $M^z_{jn,im} =(M^z_{im,jn})^*$.}
    \begin{ruledtabular}
        \begin{tabular}{r c c c}
       & $c\alpha,n$& $h\alpha,n$ & $l\alpha,n$\\
      \hline
        $c\alpha,m$ & Eq.\,(\ref{eq:Misb}) & $0$ & $\displaystyle\sqrt{\frac{2}{3}}\cos\theta \hspace{1.5pt}\delta_{nm}$\\[8pt]
      
        $c\beta,m$ & $0$ & $\displaystyle-\frac{1}{\sqrt{2}}\sin\theta \hspace{1.5pt}\delta_{nm}$ & $\displaystyle\frac{1}{\sqrt{6}}\sin\theta \hspace{1.5pt}\delta_{nm}$\\[8pt]
     
      $h\alpha,m$ & $0$ & Eq.\,(\ref{eq:Misb}) & $0$\\[8pt]
      
      $h\beta,m$ & $\displaystyle-\frac{1}{\sqrt{2}}\sin\theta \hspace{1.5pt}\delta_{nm}$ & $0$ & $0$\\[8pt]
      
      $l\alpha,m$ & $\displaystyle\sqrt{\frac{2}{3}}\cos\theta \hspace{1.5pt}\delta_{nm}$ & $0$ & Eq.\,(\ref{eq:Misb})\\[8pt]
      
      $l\beta,m$ & $\displaystyle\frac{1}{\sqrt{6}}\sin\theta \hspace{1.5pt}\delta_{nm}$ & $0$ & $0$\\
    \end{tabular}

\end{ruledtabular}
\end{table*}  
\begin{table*}[p]
    \caption{\label{tab:YMatrixElements} Optical matrix elements $M^y_{jn,im} = \hat{\mathbf{y}}\cdot\mathbf{M}_{jn,im}$. Only transitions to $\alpha$ spin states are shown since transitions to $\beta$ states can be found from the relation $M^y_{jn,im} =(M^y_{im,jn})^*$.}
    \begin{ruledtabular}
        \begin{tabular}{r c c c}
       & $c\alpha,n$& $h\alpha,n$ & $l\alpha,n$\\
      \hline
        $c\alpha,m$ & $\displaystyle\frac{m_0}{m_{c}^t}\frac{2\kappa}{\epsilon_p}\sin{\phi}\hspace{1.5pt}\delta_{nm}$ & $0$ & $\displaystyle\sqrt{\frac{2}{3}}\sin\theta\sin\phi\hspace{1.5pt}\delta_{nm}$\\
      
        $c\beta,m$ & $0$ & $0$ & $\displaystyle-\frac{1}{\sqrt{6}}(\cos\theta\sin\phi + i\cos\phi)\delta_{nm}$\\
      
      $h\alpha,m$ & $0$ & $\displaystyle\frac{m_0}{m_{h}^t}\frac{2\kappa}{\epsilon_p}\sin\phi\hspace{1.5pt}\delta_{nm}$ & $0$\\
      
      $h\beta,m$ & $\displaystyle\frac{1}{\sqrt{2}}(\cos\theta\sin\phi - i\cos\phi)\delta_{nm}$ & $0$ & $0$\\
      
      $l\alpha,m$ & $\displaystyle\sqrt{\frac{2}{3}}\sin\theta\sin\phi\hspace{1.5pt}\delta_{nm}$ & $0$ & $\displaystyle\frac{m_0}{m_{l}^t}\frac{2\kappa}{\epsilon_p}\sin\phi\hspace{1.5pt}\delta_{nm}$\\
      
      $l\beta,m$ & $\displaystyle-\frac{1}{\sqrt{6}}(\cos\theta\sin\phi + i\cos\phi)\delta_{nm}$ & $0$ & $0$\\
    \end{tabular}

\end{ruledtabular}
\end{table*}
\clearpage
The $\phi$ integration has been reduced to the average over a single term denoted with angular brackets. Performing the integration for each pair of bands gives 
\begin{align}
    \braket{|M_{\mathrm{c}\alpha,\mathrm{l}\alpha}^y \sin\phi|^2} &= \frac{1}{8}(1 - \cos{2\theta}) \nonumber \\
    \braket{|M_{\mathrm{c}\alpha,\mathrm{l}\beta}^y \sin\phi|^2} &= \frac{1}{96}(5 + 3\cos{2\theta}) \nonumber \\
    \braket{|M_{\mathrm{c}\alpha,\mathrm{h}\beta}^y \sin\phi|^2} &= \frac{1}{96}(17 - 9\cos{2\theta}).
\end{align}

Note that if we chose $x$ polarized light, we would need to use the $x$ components of the interband matrix elements and take $\hat{\mathbf{e}}\cdot\mathbf{k}_t = \cos\phi$. The result is that the integration over $\phi$ yields identical values to $y$ polarized light. This equivalence is consistent with the fact that physical measurements must have the azimuthal symmetry of the isotropic bands.

\section{Results}\label{sec:Results}

Normalized transmission versus pump-probe delay was measured as described in Sec.\,\ref{sec:Experiment}, and coupled nonlinear Schr{\"o}dinger equations were used to fit the curves to ND-2PA coefficients. The exact procedure is detailed in Appendix \ref{sec:Propagation}, along with all necessary approximations and empirical adjustments.

\begin{figure}[htb]
    \centering
    \includegraphics{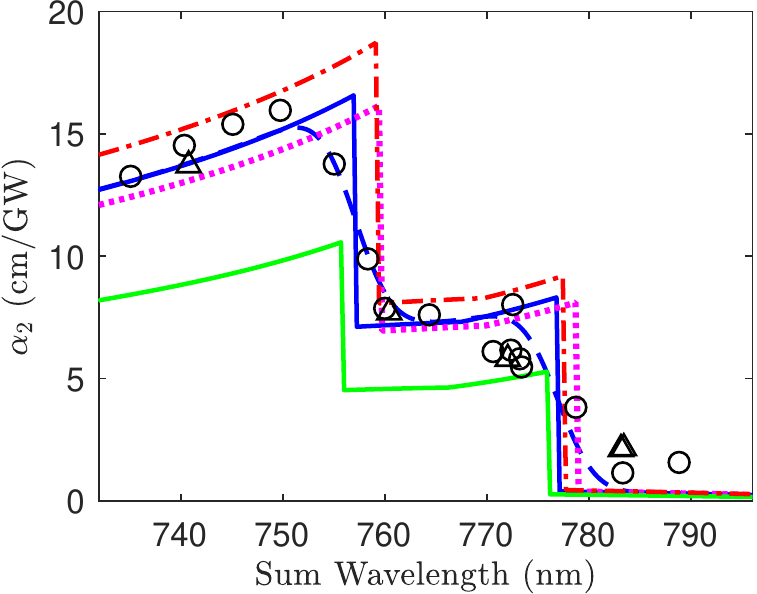}
    \caption{TM-TM ND-2PA coefficients vs $\lambda_{\mathrm{sum}} = (1960^{-1} + \lambda_1^{-1})^{-1}$ (nm). Black circles: measured coefficients with total pulse energy $E_2 = 3.5 \,\mathrm{pJ}$. Black triangles: measured coefficients with $E_2 = 4.9 \,\mathrm{pJ}$. Red dash-dotted line: theory with 8 band basis. Solid blue line: theory with 14 band basis. Dashed blue line: TM-TM theory convolved with $4 \,\mathrm{nm}$ Gaussian to approximate bandwidth effects. Magenta dotted line: 14 band model without adjustments to well width and barriers. Solid green line: 14 band calculation with $E_p = 25$\,eV}.
    \label{fig:2PATMCurves}
\end{figure}

TM-TM measurement results are plotted versus sum wavelength in Fig.\,\ref{fig:2PATMCurves} alongside theoretical predictions. Sum wavelength is defined by $1/\lambda_{\mathrm{sum}} = 1/\lambda_1 + 1/\lambda_2$, where the pump wavelength $\lambda_2$ is fixed at 1960\,nm. Excitations of light hole states bring about discontinuities in the spectrum, with the $l1\rightarrow c2$ and $l2 \rightarrow c1$ transitions accounting for the shorter and longer wavelength steps, respectively. In contrast with light hole contributions, heavy hole signals exhibit a gradual increase due to the interband $y$ matrix element's $\sin \theta$ (forbidden) dependence. Each transition's onset is found by subtracting the subband energies in Table \ref{tab:QWEnergies} of Appendix \ref{sec:Material} using the selection rule $m - n = \pm 1$.

Expanding states in the 8 band basis with $E_p = 28.9$\,eV \cite{Hermann1977}, our predicted curve matched the data apart from a $3$\,nm wavelength shift. This offset is mitigated by using the 14 band model with $E_p' = 6 \,\mathrm{eV}$ \cite{Hermann1977}. In both cases, we assumed there were small growth errors such that the real material consisted of $7.84$\,nm wells with Al$_{0.328}$Ga$_{0.672}$As barriers. The effect of this modification is revealed by comparing the blue (solid) and magenta (dotted) curves of Fig.\,\ref{fig:2PATMCurves}. Other sources for this wavelength inaccuracy could be OSA miscalibration or the use of imprecise bandgap values in simulations, but these assumptions lead to curves nearly identical to the ones shown.

Interestingly, calculated 2PA coefficients vary significantly with the value chosen for the Kane energy. This sensitivity is apparent when comparing the 14 band theoretical curve with another that has $E_p = 25 \,\mathrm{eV}$ and $E_p' = 6 \,\mathrm{eV}$ \cite{Hermann1977}. By the arguments of Sec.\,\ref{ssec:Envelopes}, under these conditions we require $m_0/\mathscr{M}_{cc}^{zz} = 3.0$ so that Eq.\,(\ref{eq:mc2}) reduces to the bulk effective mass for GaAs. For comparison, $m_0/\mathscr{M}_{cc}^{zz} = -2.2$ when $E_p = 28.9$\,eV. This modification reduces the intersubband matrix element according to Eq.\,(\ref{eq:Misb}).

\begin{figure}[!tb]
    \centering
    \includegraphics{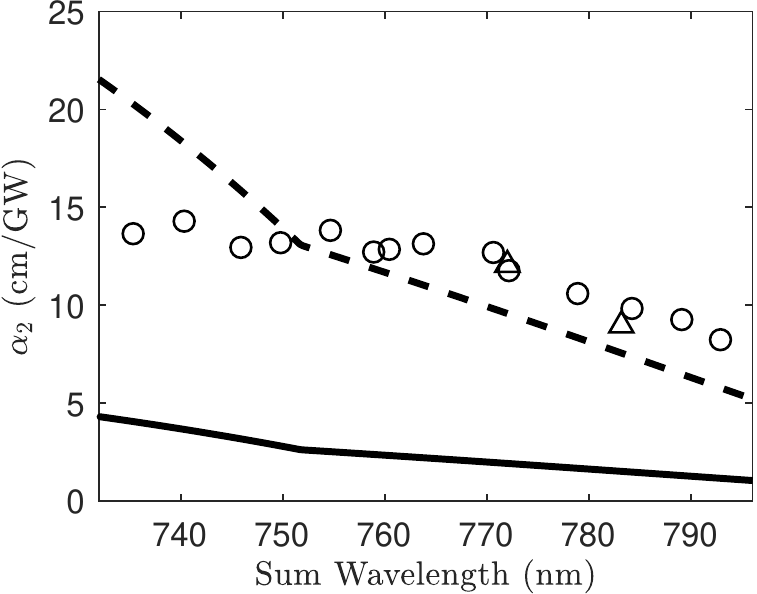}
    \caption{Comparison between theory and experiment for TE-TE polarization. Black circle TE-TE ND-2PA coefficients at $E = 4.2 \,\mathrm{pJ}$. Black diamond: TE-TE ND-2PA coefficients with $E = 6.0 \,\mathrm{pJ}$. Black line: TE-TE 2PA theory. Black dotted line: TE-TE 2PA theory multiplied by $4$.}
    \label{fig:2PATECurves}
\end{figure}

Fig.\,\ref{fig:2PATECurves} shows the TE-TE results compared with 14 band calculations. The ND-2PA edge is energetically lower than in TM-TM because it first occurs for $h1\rightarrow c1$ transitions, leading to large anisotropy below the $l1 \rightarrow c2$ TM-TM transition energy. The theoretical curve, which is smooth with a kink at $750\,\mu$m from $h2\rightarrow c2$ transitions, shows this anisotropy. However, our unscaled calculations differ from the measurements by over a factor of four and show some dissimilarity in shape. We provide reasons for these differences in the following section. 

\section{Discussion}\label{sec:Discussion}
The TM-TM theory matches the data without any non-physical scaling parameter; this excellent agreement is likely due to the dominance of allowed transitions. The parabolic band approximation works well because important features in the 2PA spectrum occur at small $\mathbf{k}_t$, where the parabolic band approximation introduces little error. Calculations show that the ND-2PA coefficients measured here are enhanced by a factor of $1.54$ over degenerate 2PA.

We do not notice any bound excitonic response, which may be attributed to temperature effects. Continuum exciton enhancement is also not evident. Ref. \cite{Shimizu1989} concludes that this contribution is absent in Ref. \cite{Tai1989} due to low sample quality and large exciton spatial extent. We suspect the same reasons apply here, with further reductions possibly occurring due to loss of 2-D character from interactions between many closely spaced wells.

In contrast to TM-TM polarizations, the parabolic band approximation introduces significant errors in TE-TE ND-2PA coefficients. By ignoring unit cell intermixing, we underestimate the $\mathbf{k}_t$-dependent scaling of forbidden transitions that are necessary in TE-TE pathways. We also determine that it is insufficient to examine only self-transitions as the forbidden step; we must also consider intersubband transitions and those between different hole types. Away from the band edge, light hole to heavy hole transitions were shown to be non-negligible for 2PA in bulk semiconductors \cite{Hutchings1992c}. TE-TE coefficients could be more accurately calculated by numerically computing the highly non-parabolic band dispersion as in Ref. \cite{Chuang1991}. Eq.\,(\ref{eq:normalizedM}) would then give matrix elements throughout the Brillouin zone, which are used to find 2PA coefficients according to Eq.\,(\ref{eq:SpectralFunction}).

The sensitivity of the 2PA coefficient to Kane energy and effective masses indicates that pump probe spectroscopy of quantum wells may be an effective method for determining basic material parameters. Our results, while not precise enough to justify a definitive declaration, seem to support the idea that the Kane energy is closer to $28.9$\,eV than lower values that have been reported (See Appendix \ref{sec:Material}). Furthermore, \citet{Hubner2009} and others have shown evidence that the Kane energy is dependent on temperature. With better spectral resolution and careful experimental setup, the temperature dependence of $E_p$ could be reflected both in a magnitude change of the normalized transmission signal and a shift of the ND-2PA edge as effective mass is changed (See Eq.\,(\ref{eq:EffectiveMassScaling})). 

\begin{figure}[htb]
    \centering
    \includegraphics{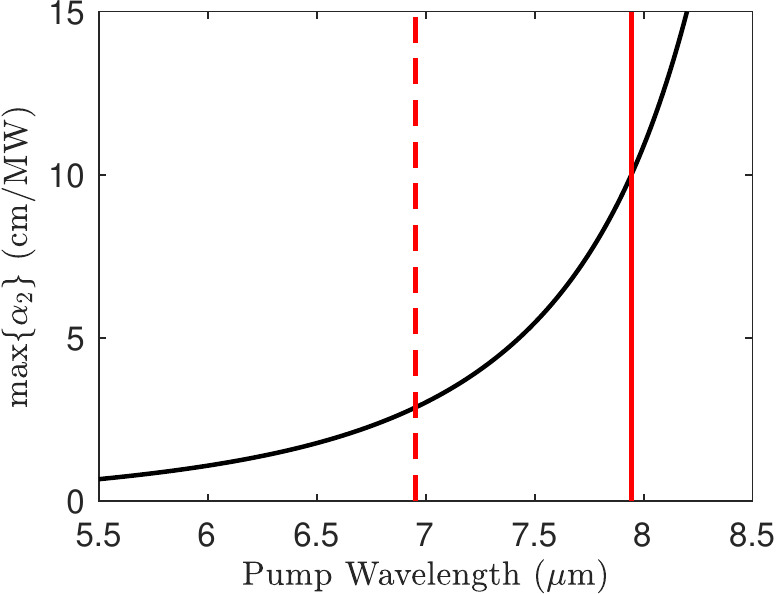}
    \caption{Maximum TM-TM ND-2PA coefficient  (in cm/MW) vs. pump wavelength. Red dashed line denotes point where maximum 2PA occurs for a probe wavelength within $25$\,meV of the $h1 \rightarrow c1$ bandgap. Red solid line is the same for the $l1 \rightarrow c1$ gap.}
    \label{fig:2PAEnhancement}
\end{figure}
With this experimental support for our model, we can re-examine ND-2PA for extremely nondegenerate photon pairs \cite{Pattanaik2016, Cirloganu2011a, Pasquarello1990a}. Fig.\,\ref{fig:2PAEnhancement} shows that the calculated TM-TM ND-2PA coefficients rapidly increase as the pump wavelength nears the $c1 \rightarrow c2$ resonance at $9.55\,\mu$m. The red vertical lines denote points where the probe energy for maximum 2PA lies within $kT = 25$\,meV of the forbidden (dashed, $h1 \rightarrow c1$) and allowed (solid, $l1 \rightarrow c1$) linear absorption edges. The offset value is chosen so that we can roughly assume negligible impurity state absorption, but these edges will shift depending on material quality and temperature. For $\lambda_2 = 7.5\,\mu$m, we see that $\mathrm{max}\{\alpha_2\} = 5.7$\,cm/MW---an enhancement of  $\sim 360\times$ over the slightly nondegenerate case studied here, and a considerably larger ND-2PA coefficient than any we have measured in bulk semiconductors ($\sim 1$cm/MW) \cite{Cirloganu2011a}. This enhancement suggests the possibility of extremely sensitive gated detection as in Ref.\,\cite{Fishman2011}, with even further increases in photogenerated carrier density due to long pulse interactions within a waveguide.

\acknowledgments
Funding for NC, DH, and EVS was provided by National Science Foundation grant DMR-1609895 and the Army Research Laboratory (W911NF-15-2-0090). This work was performed, in part, at the Center for Integrated Nanotechnologies, an Office of Science User Facility operated for the U.S. Department of Energy (DOE) Office of Science by Los Alamos National Laboratory (Contract 89233218CNA000001) and Sandia National Laboratories (Contract DE-NA-0003525). JW and SPG acknowledge the support from the Fonds
de la Recherche Fondamentalei Collective (FRFC) (PDR.T.1084.15).
\appendix
\section{Kane Band Structure}\label{sec:BulkBandStructure}
\citet{Kane1957} developed a band structure model for bulk zinc blende materials using a $\mathbf{k}\cdot \mathbf{p}$ formalism including spin-orbit interaction. The unit cell basis consisted of two spin-degenerate $S$-like functions $\ket{S\uparrow}$ and $\ket{S\downarrow}$ with energy $E_g$ and six degenerate $P$-like functions $\ket{X\uparrow}$, $\ket{X\downarrow}$, $\ket{Y\uparrow}$, $\ket{Y\downarrow}$, $\ket{Z\uparrow}$, and $\ket{Z\downarrow}$ with $E=0$. By symmetry of the zincblende crystal, all non-zero momentum matrix elements are given by \cite{Yu1980}
\begin{equation}\label{eq:P}
P = \frac{\hbar}{m_0}\braket{iS|p_x|X} = \frac{\hbar}{m_0}\braket{iS|p_y|Y}= \frac{\hbar}{m_0}\braket{iS|p_z|Z}.
\end{equation}
The wave vector in bulk materials is not restricted to two dimensions as in quantum wells. In Ref.\,\cite{Kane1957}, the $\mathbf{k}\cdot \mathbf{p}$ Hamiltonian is diagonalized in a rotated coordinate system for which $\mathbf{k} = k\hat{\mathbf{z}}$. This coordinate transformation is represented as a 3-dimensional rotation matrix since $X$, $Y$ and $Z$ transform as the components of a vector \cite{Yu1980}. Finally, the k-dependent eigenstates are found to be
\begin{align}\label{eq:DiagHam}
    u_{h\alpha} &= -\sqrt{1/2}\Ket{(X+iY)\uparrow}' \nonumber \\
    u_{h\beta} &= \sqrt{1/2}\Ket{(X-iY)\downarrow}' \nonumber \\
    u_{j\alpha} &= a_j\Ket{iS\downarrow}' + b_j/\sqrt{2}\Ket{(X-iY)\uparrow}'+ c_j\Ket{Z\downarrow}' \nonumber \\
    u_{j\beta} &= a_j\Ket{iS\uparrow}' + b_j/\sqrt{2}\Ket{-(X+iY)\downarrow}'+ c_j\Ket{Z\uparrow}'.
\end{align}
The heavy hole bands ($h$) are uncoupled while the conduction, light hole, and split off bands---denoted by index $j$---intermix. The primed kets indicate rotated basis functions. 

The zone center ($\mathbf{k} = 0$) unit cell functions are listed in Table \ref{tab:Band}. Taking the first four  $\alpha$ and $\beta$ states gives the 8 band Kane basis described above. The more complete 14 band model includes conduction bands at energy $E_g'$ and $E_{s'} = E_g' - \Delta'$, with $\Delta'$ representing spin-orbit splitting in the conduction bands. These two sets are used as bases for the envelope expansion in Eq.\,(\ref{eq:envExp}).
\begin{table}[htb]
    \caption{Table of zone center wave functions and their corresponding energies.}
    \begin{ruledtabular}\label{tab:Band}
        \begin{tabular}{cl}
        E & $u_{j0}(\mathbf{r})$ \\
        \hline \\[-7pt]
        $E_g$ & $u_{c\alpha}: \Ket{iS\downarrow}$\\
        $0$ & $u_{h\alpha}:-\sqrt{1/2}\Ket{(X+iY)\uparrow}$\\
        $0$ & $u_{l\alpha}:\sqrt{1/6}\Ket{(X-iY)\uparrow} + \sqrt{2/3}\Ket{Z\downarrow}$\\
        $-\Delta$ & $u_{s\alpha}:\sqrt{1/3}\Ket{(X-iY)\uparrow} - \sqrt{1/3}\Ket{Z\downarrow}$\\
        $E_g'$ & $u_{h'\alpha}:-\sqrt{1/2}\Ket{(X^c+iY^c)\uparrow}$\\
        $E_g'$ & $u_{l'\alpha}:\sqrt{1/6}\Ket{(X^c-iY^c)\uparrow} + \sqrt{2/3}\Ket{Z^c\downarrow}$\\
        $E_{s'}$ & $u_{s'\alpha}:\sqrt{1/3}\Ket{(X-iY)\uparrow}-\sqrt{1/3}\Ket{Z^c\downarrow}$\\[4pt]
        \hline \\[-7pt]
        $E_g$ & $u_{c\beta}:\Ket{iS\uparrow}$\\
        $0$ & $u_{h\beta}:\sqrt{1/2}\Ket{(X-iY)\downarrow}$\\
        $0$ & $u_{l\beta}:-\sqrt{1/6}\Ket{(X+iY)\downarrow} + \sqrt{2/3}\Ket{Z\uparrow}$\\
        $-\Delta$ & $u_{s\beta}:\sqrt{1/3}\Ket{(X+iY)\downarrow}+ \sqrt{1/3}\Ket{Z\uparrow}$\\
        $E_g'$ & $u_{h'\beta}:\sqrt{1/2}\Ket{(X^c-iY^c)\downarrow}$\\
        $E_g'$ & $u_{l'\beta}:-\sqrt{1/6}\Ket{(X^c+iY^c)\downarrow} + \sqrt{2/3}\Ket{Z^c\uparrow}$ \\
        $E_{s'}$ & $u_{s'\beta}:\sqrt{1/3}\Ket{(X^c+iY^c)\downarrow}+ \sqrt{1/3}\Ket{Z^c\uparrow}$\\[3pt]
\end{tabular}
\end{ruledtabular}
\end{table}
\subsection*{Interband matrix elements}\label{subsec:InterbandElement}
Comparing the $u_{j\mathbf{k}}(\mathbf{r})$ of Eq.\,(\ref{eq:DiagHam}) to the zone center $u_{j0}(\mathbf{r})$ values in Table \ref{tab:Band}, we note they differ by the expansion coefficients as well as a rotation of basis functions. As in Ref.\,\cite{Yamanishi1984}, we assume quantum well transitions are adequately described by using the zone center expansion coefficient while applying the basis rotation. For example, $\braket{u_{l\alpha}|p_z|u_{c\alpha}}' = \braket{iS|p_z|\sqrt{2/3}(Z \cos\theta)}$, which evaluates to the value given in Table \ref{tab:ZMatrixElements} after application of Eqs.\,(\ref{eq:P}) and (\ref{eq:interband}).
\section{Intersubband Matrix Element}\label{sec:IntersubbandElement}
This appendix derives the intersubband matrix element in Eq.\,(\ref{eq:Misb}). The procedure shown is for conduction bands in the 8 band basis, but it is easily generalized to other bands and different basis sets. 

The momentum matrix element between states given in the envelope expansion (Eq.\,(\ref{eq:envExp})) is
\begin{equation}
    P_{jn,im}^z = \braket{\psi_{jn},p_z\psi_{im}} = \sum_{\mu \nu}\Braket{u_{\mu} \chi^\mu_{jn}, p_z u_{\nu}\chi^\nu_{im}},
\end{equation}
where we have chosen to represent inner products with operator $\hat{T}$ as $\braket{f,\hat{T}g}$ for clarity. As usual, $\braket{\lambda f, g} = \lambda^* \braket{f,g}$. We apply the chain rule for the differential operator $p_z$, then assume $\chi$ and $u$ change on different enough scales so that we can integrate their expressions separately. In atomic units we find
\begin{align}\label{eq:isbexp}
    P_{jn,im}^z &= \sum_{\mu \nu} \Braket{\chi^{\mu}_{jn}, p_z\chi^{\nu}_{im}}\delta_{\mu\nu} + p_{\mu\nu}^z\Braket{\chi^{\mu}_{jn}, \chi^{\nu}_{im}} \nonumber \\
                                &= \braket{\chi^c_{jn}, p_z \chi^c_{im}} + \braket{\chi^l_{jn}, p_z \chi^l_{im}} + \braket{\chi^s_{jn}, p_z \chi^s_{im}} \nonumber \\
                                &+ \sqrt{\frac{2}{3}}P \braket{\chi^l_{jn}, \chi^c_{im}} -\sqrt{\frac{1}{3}}P\braket{\chi^s_{jn}, \chi^c_{im}}  \nonumber \\
                                &+ \sqrt{\frac{2}{3}}P \braket{\chi^c_{jn}, \chi^l_{im}} - \sqrt{\frac{1}{3}}P \braket{\chi^c_{jn}, \chi^s_{im}}, 
\end{align}
where $p_{\mu\nu}^z$ values are obtained from Eq.\,(\ref{eq:P}) and Table \ref{tab:Band}. Suppose now that the intial and final envelopes are conduction states. We can express the non-dominant envelopes as a function of the conduction envelope by
\begin{align}\label{eq:chisub}
    \chi^l &= \sqrt{\frac{2}{3}}P\frac{1}{E - V_v(z)}p_z\chi^c \nonumber \\
    \chi^s &= -\sqrt{\frac{1}{3}}P\frac{1}{E + \Delta - V_v(z)}p_z\chi^c. 
\end{align}
These substitions appear in Ref. \cite{Bastard1976}, and are re-derived in detail in the supplemental material \cite{Supp}. Applying Eqs.\,(\ref{eq:chisub}) to Eq.\,(\ref{eq:isbexp}) gives
\begin{align}
    P_{cn,cm}^z &= \frac{1}{2}\left[\Braket{p_z\chi^c_{n}, \chi^c_{m}} + \frac{4P^2}{3} \Braket{\frac{1}{E_{n} - V_v(z)}p_z\chi^c_{n}, \chi^c_{m}} \right. \nonumber \\ 
                                  &\hspace{45pt}+ \left. \frac{2P^2}{3}\Braket{\frac{1}{E_{n} + \Delta - V_v(z)}p_z\chi^c_{n}, \chi^c_{m}}\right] \nonumber \\
                                    &+ \frac{1}{2}\left[\Braket{\chi^c_{n}, p_z \chi^c_{m}} + \frac{4P^2}{3} \Braket{\chi^c_{n}, \frac{1}{E_{m} - V_v(z)}p_z\chi^c_{m}} \right. \nonumber \\
                                  &\hspace{45pt}+ \left. \frac{2P^2}{3} \Braket{\chi^c_{n}, \frac{1}{E_{m} + \Delta - V_v(z)}p_z\chi^c_{m}}\right],
\end{align}
where we dropped the subscript $c$ on the right hand side. We also used the fact that $\braket{\chi_{n}^c, p_z\chi_{m}^c} = [\Braket{p_z\chi^c_{n}, \chi^c_{m}} + \Braket{\chi^c_{n}, p_z\chi^c_{m}}]/2$. Noting that $E_p = 2P^2$ in atomic units, comparing to Eq.\,(\ref{eq:mc}) immediately leads to
\begin{align}
    \braket{\psi_{cn}, p_z \psi_{cm}} &= \frac{1}{2}\Braket{\left[\frac{1}{m(E_n, z)} - \frac{1}{\mathscr{M}_{cc}^{zz}}\right]p_z\chi^c_{n},  \chi^c_{m}} \nonumber \\
                                &+ \frac{1}{2}\Braket{\chi^c_{n}, \left[\frac{1}{m(E_m, z)} - \frac{1}{\mathscr{M}_{cc}^{zz}}\right] p_z \chi^c_{m}}. 
\end{align}
Normalization and conversion back to bra-ket notation gives Eq.\,(\ref{eq:Misb}). This process can be easily repeated for hole transitions and states written in the 14 band basis.
\section{Simulation parameters and procedures}\label{sec:Material}

Since we are not working with an idealized structure, we must employ some numerical techniques to model our systems. In the first subsection of this appendix, we calculate energy levels and envelopes for states in the finite quantum well. In the second, we find the optical mode structure and dispersion characteristics of the waveguide.
\subsection{Material simulations}
We first give the bandgap of GaAs and the composition-dependent bandgap of AlGaAs in order to determine the confining potential imposed by the AlGaAs barriers. Then we provide values for interband couplings and effective masses, followed by a brief summary of the calculation results. Each value is taken from literature, making adjustments as needed.

The temperature-dependent bandgap of GaAs is given by the Varshni relation \cite{Varshni1967}
\begin{equation}\label{eq:bg}
    E_g = 1.519 -\alpha \frac{T^2}{T + \beta} \hspace{15pt} [\mathrm{eV}]
\end{equation}
with $\alpha = 8.95\times 10^{-4} \,\mathrm{eV/K}$ and $\beta = 538 \,\mathrm{K}$ \cite{ElAllali1993,Vurgaftman2001}. The higher conduction band energies in the 14 band model are taken to be $E_g' = 4.63$\,eV and $E_{s'} = 4.44$\,eV\cite{Cardona1967}.

The potential barriers imposed by the quantum well layer structure come from the empirical expression for total band offset between GaAs and Al$_x$Ga$_{1-x}$As:\cite{ElAllali1993}
\begin{equation}\label{eq:compscale}
    \Delta E_g = 1.395x \quad (x \leq 0.41).
\end{equation}
The ratio $Q = \Delta E_v/\Delta E_g = 0.33$ \cite{Vurgaftman2001} at an Al$_x$Ga$_{1-x}$As interface, simplifying conduction and valence offsets to 
\begin{align}\label{eq:compdep}
    \Delta E_c &= V_c = 0.963x \quad \mathrm{[eV]} \nonumber \\
    \Delta E_v &= V_{h} = -0.432x \quad \mathrm{[eV]}.
\end{align}
All hole types are presumed to have offset $V_h$ and all conduction bands are taken to have offset $V_c$. 

Literature values for the Kane energy of GaAs vary between $E_p = 22.9 \,\mathrm{eV}$\cite{Balslev1969}, $25.7 \,\mathrm{eV}$ \cite{Lawaetz1971}, $27.86 \,\mathrm{eV}$ \cite{Pfeffer1996}, $28.8 \,\mathrm{eV}$\cite{Eppenga1987} and $28.9 \,\mathrm{eV}$ \cite{Hermann1977}. We choose $E_p = 28.9 \,\mathrm{eV}$, and take the inter-conduction band coupling strength as $E_p' = 6\,\mathrm{eV}$ \cite{Hermann1977}. 

The conduction band effective mass of GaAs is $m_{c}^z = m_c^t = 0.0635 m_0$ \cite{Vurgaftman2001} at room temperature. Light hole masses are anisotropic with $m^z_{l} = -0.082 m_0$ \cite{Vrehen1968} as determined from cyclotron resonance at $77 \,\mathrm{K}$. This mass is adjusted to its room temperature value by
\begin{equation}\label{eq:EffectiveMassScaling}
    \frac{1}{m_{l, 295\,\mathrm{K}}} = \frac{1}{m_{l,77\,\mathrm{K}}} + \frac{2E_p}{3} \left(\frac{1}{E_{g,295\,\mathrm{K}}} - \frac{1}{E_{g,77\,\mathrm{K}}}\right),
\end{equation}
where temperature-dependent bandgaps are taken from Eq.\,(\ref{eq:bg}). This relation comes from subtracting the Eq.\,(\ref{eq:mc}) expressions for $295$\,K and $77$\,K. The final outcome is that $m_{l, 295\,\mathrm{K}}^z = -0.077m_0$. 

The energy-dependent effective mass is calculated in the 8 band model using Eq.\,(\ref{eq:mc}) with spin-orbit splitting of $\Delta = 0.341 \,\mathrm{eV}$ \cite{Aspnes1973}. 

The heavy hole effective mass $m_h(z)$ is \cite{Chuang1995}
\begin{align}\label{eq:LuttingerZ}
    m_{h}^z(z) = \frac{m_0}{\gamma_1(x(z)) - 2\gamma_2(x(z))}, 
\end{align}
where $x(z)$ is the AlGaAs composition at position $z$. The symbol $\gamma(x)$ is the six-band Luttinger parameter linearly interpolated between the values for GaAs and AlAs \cite{Vurgaftman2001}:
\begin{align}
    \gamma_1(x) &= 6.98 - (3.76 - 6.98)x \nonumber \\
    \gamma_2(x) &= 2.06 - (0.82 - 2.06)x.
\end{align}
Transverse hole effective masses are also taken from the Luttinger parameters as \cite{Chuang1995}
\begin{align}\label{eq:LuttingerMasses}
    m_{h}^t &= \frac{m_0}{\gamma_1(0) + \gamma_2(0)} \\
    m_{l}^t &= \frac{m_0}{\gamma_1(0) - \gamma_2(0)}. 
\end{align}

With all material parameters known, the shooting method \cite{Harrison2005} is used to solve Equation $(\ref{eq:SEM})$. The procedure yields the energy level (See Table \ref{tab:QWEnergies}) and dominant wavefunction envelope for each state. Note that the material widths and compositions are slightly altered as explained in Sec.\,\ref{sec:Results}.
\begin{table}[htb]
    \caption{Quantum well subband energies (in meV) at $\kappa = 0$ for $7.84\,\mathrm{nm}$ GaAs quantum wells with Al$_{0.328}$Ga$_{0.672}$As barriers.}
    \label{tab:QWEnergies}
    \begin{ruledtabular}
    \begin{tabular}{rccc}
      & c & lh & hh \\ 
      \hline
    $1$ & $1475.9$ & $-33.5$ & $-11.5$ \\
    $2$ & $1604.9$ &  $-120.3$ & $-45.4$ \\
    $3$ & -- & -- & $-98.9$ \\
\end{tabular}
\end{ruledtabular}
\end{table}
\subsection{Waveguide Modes}
Refractive index values for the various AlGaAs compositions used were calculated by Adachi’s formulas \cite{Adachi1989}. The index of the quantum well active region was estimated to be the spatial average of the well ($w$) and barrier ($b$) permittivities
\begin{equation}\label{eq:AveragedIndex}
    n^2(x,\lambda)= \frac{L_{w} n_{w}(x,\lambda)^2+L_b n_b(x,\lambda)^2}{L_{w}+L_b}.
\end{equation}

\begin{figure}[tb]
     \centering
     \includegraphics{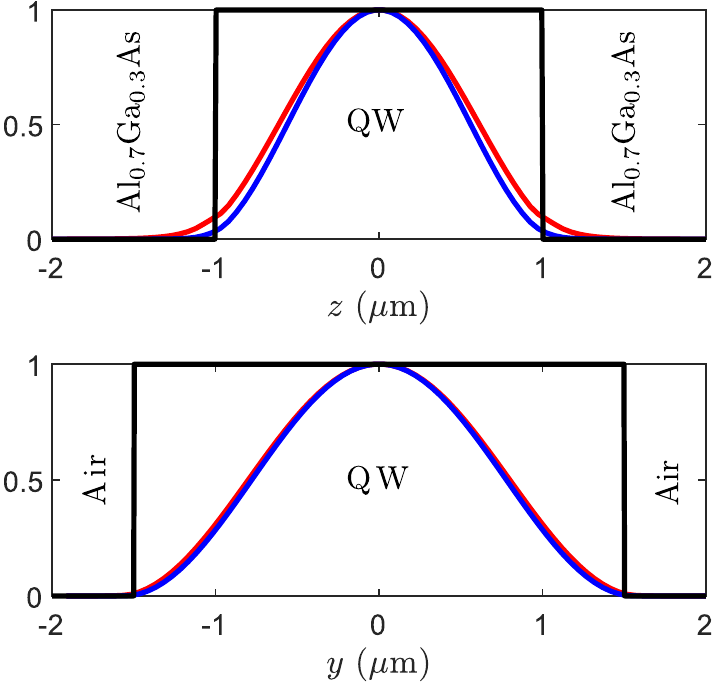}
     \caption{Normalized pump (red) and probe (blue) TM mode profiles for the structure shown in Fig.\,\ref{fig:Sample}.}
     \label{fig:ModeAreas}
 \end{figure}

Using these indices, electromagnetic mode profiles and dispersion curves were calculated from the finite difference method with Lumerical MODE Solutions \cite{Lumerical}. The mode shapes are shown in Fig.\,\ref{fig:ModeAreas}.
 The third order mode areas $a_i$ and overlap $\eta_{ij}$ were calculated in the usual way \cite{Lin2007a}:
 \begin{equation}
     a_i = \frac{\left[\iint |F_i(y,z)|^2 dy dz\right]^2}{\iint |F_i(y,z)|^4 dy dz},
 \end{equation}
 \begin{equation}
     \eta_{ij} = \frac{\iint |F_i(y,z)|^2|F_j(y,z)|^2 dy dz}{\left[\iint |F_i(y,z)|^4 dy dz\right]^{1/2}\left[\iint |F_j(y,z)|^4 dy dz\right]^{1/2}},
 \end{equation}
 where $F_i(y,z)$ is the electric field profile of mode $i$. The calculated TM (TE) mode area at $1960 \,\mathrm{nm}$ was found to be $3.56$ ($3.48$)$\,\mu\mathrm{m}^2$, and the TM (TE) probe mode areas range from $3.14$ ($3.07$)$\,\mu\mathrm{m}^2$ at $1150 \,\mathrm{nm}$ to $3.21$ ($3.17$)$\,\mu\mathrm{m}^2$ at $1350 \,\mathrm{nm}$. The mode overlap at $1176 \,\mathrm{nm}$ and $1960 \,\mathrm{nm}$ was $\eta_{12} = 0.9967$, so they are treated as unity. Due to tight optical confinement within the active region, propagation is well-approximated by taking modes to travel through a material entirely made up of the quantum wells.

 We numerically differentiated the refractive index curves to find group velocities and second order dispersions. The largest GVM of $\rho = \Delta \beta = 860$\,fs/mm occurred when $\lambda_1 = 1176 \,\mathrm{nm}$ and $\lambda_2 = 1960\,\mathrm{nm}$, with the pump travelling faster than the probe. Around this value, $150$\,fs pulses walk off from each other on the length scale of $350\,\mathrm{\mu m}$. The largest dispersion coefficient of $\beta_2 = 1841\,\mathrm{fs^2/mm}$ occurs at the same probe wavelength, while the pump dispersion is $\beta_2^{(2)} = 741 \,\mathrm{fs^2/mm}$ at $1960$\,nm. These values were used in the simulations of Appendix \ref{sec:Propagation}.
\section{Nonlinear propagation and data fitting}\label{sec:Propagation}
We modeled pulse propagation by the coupled nonlinear Schrodinger equations \cite{Agrawal2012}
\begin{align}\label{eq:NLS1}
    \left(\vphantom{\frac{\beta_2^{(2)}}{2}}\frac{\partial{}}{\partial{x}}  + \beta_1^{(1)}\frac{\partial{}}{\partial{t}} +\right.&\left. i\frac{\beta_2^{(1)}}{2}\frac{\partial^2{}}{\partial{t^2}}\hphantom{i}+ \frac{\sigma_1}{2}\right)A_1 \nonumber \\
    &= i\left(\gamma_{11}|A_1|^2 + 2\gamma_{12}|A_2|^2 \right)A_1,
\end{align}
\vspace{-15pt}
\begin{align}\label{eq:NLS2}
    \left(\vphantom{\frac{\beta_2^{(2)}}{2}}\frac{\partial{}}{\partial{x}}  + \beta_1^{(2)}\frac{\partial{}}{\partial{t}} +\right.&\left. i\frac{\beta_2^{(2)}}{2}\frac{\partial^2{}}{\partial{t^2}}\hphantom{i}+ \frac{\sigma_2}{2}\right)A_2 \nonumber \\
    &= i\left(\gamma_{22}|A_2|^2 + 2\gamma_{21}|A_1|^2 \right)A_2.
\end{align}
$|A_{l}|^2$ is the instantaneous power, $\beta_1^{(l)}$ and $\beta_2^{(l)}$ are the first and second order dispersion, and $\sigma_l$ is the loss. We neglect free carrier contributions to the nonlinear propagation because the pulses have sufficiently low average power such that excited carrier density is negligible. The nonlinear parameter $\gamma_{ij}$ is written in terms of the mode areas and overlap as \cite{Lin2007a}
\begin{equation}
    \gamma_{i j} = \frac{\omega_i}{c}\frac{n_2(\omega_i;\omega_j) \eta_{ij}}{\sqrt{a_i a_j}} + i\frac{\alpha_2(\omega_i;\omega_j) \eta_{ij}}{2\sqrt{a_i a_j}}.
\end{equation}
We set $\mathrm{Im}\{\gamma_{22}\} = 0$ because the pump wavelength is below the D-2PA edge, and the $\gamma_{i1}$ are ignored because the probe power is low. All nonlinear refraction effects from $\mathrm{Re}\{\gamma_{ij}\}$ are ignored, which is justified in the following subsection.
\subsection*{Raw data and analysis}
\begin{figure}[htb]
    \centering
    \includegraphics{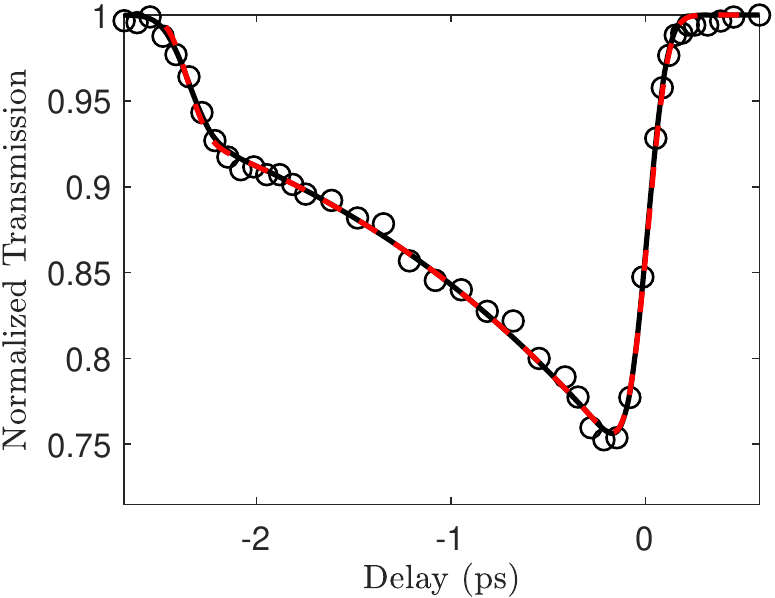}
    \caption{Normalized transmission vs delay at $\lambda_2 = 1960 \,\mathrm{nm}$ and $\lambda_1 = 1246 \,\mathrm{nm}$. The results are compared to calculations using the split-step Fourier method (black line) and analytical expression of Eq.\,(\ref{eq:analytical}) (red dashed line). The parameters used were: $\mathrm{Im}\{\gamma_{12}\} = 12.9 \,\mathrm{fs/(mm\,pJ)}$ , $E_2 = 4.8 \,\mathrm{pJ}$, $\rho = 654 \,\mathrm{fs/mm}$, $L = 3.6 \,\mathrm{mm}$, $\tau_1 = 121 \,\mathrm{fs}$, $\tau_2 = 154 \,\mathrm{fs}$, $\alpha_2 = 0.4\,\mathrm{mm}^{-1}$. The total width of the signal is $\tau = \rho L = 2.35$\,ps.}
    \label{fig:FullScanGVM}
\end{figure}

Fig.\,\ref{fig:FullScanGVM} shows a normalized transmission signal generated as described in Sec.\,\ref{sec:Experiment}. A delay of zero indicates that the pump and probe arrive at the front facet at the same time, and a negative delay means the probe arrives before the pump. The curve is temporally wider than the input pulses because the faster moving pump overtakes the probe within the sample for delays up to about $-2.3$\,ps. 

Transmission curves in high GVM experiments usually exhibit a table top shape like those in Ref\,\cite{Cirloganu2011a}. Instead, we see a decay in the nonlinear transmission magnitude as delay becomes more negative. This indicates that loss in pump irradiance leads to a smaller signal as the pulses meet after propagating farther through the sample. If these losses were caused by an interband absorption process, there would be a constant signal at positive delay due to excited carrier interactions. Instead, we attribute this decay to high scattering from roughness of the etched sidewall.

The shape of Fig.\,\ref{fig:FullScanGVM} is explained by the analytical expression for normalized transmission \cite{Supp}
\begin{align}\label{eq:analytical}
    \Delta T &= \exp\left\{ 2\frac{E_2\,\mathrm{Im}\{\gamma_{12}\}}{\rho}\exp{\left[\frac{\sigma_2 \tau}{\rho} + \left(\frac{\sigma_2\tau_x}{2\rho}\right)^2\right]} \right. \nonumber \\
             &\times \left.\left[\mathrm{erf}\left(\frac{\tau}{\tau_x} + \frac{\sigma_2 \tau_x}{2\rho}\right) - \mathrm{erf}\left(\frac{\tau + \rho L}{\tau_x} + \frac{\sigma_2 \tau_x}{2\rho}\right)\right]\vphantom{\left(\frac{\alpha_2\tau_x}{2\rho}\right)^2}\right\}.
\end{align}
Here, $L$ is the length of the sample and $E_2$ is the total energy of the pump pulse. The value $\tau_x = (\tau_1^2 + \tau_2^2)^{1/2}$ is the pure cross correlation width between pulse durations $\tau_j$, and $\rho = \beta_1^{(2)} - \beta_1^{(1)}$ is the GVM parameter. Eq.\,(\ref{eq:analytical}) was derived by ignoring second order dispersion. Despite this approximation, we see by comparing the curves in Fig.\,\ref{fig:FullScanGVM} that the equation yields nearly identical results to those found by split-step Fourier integration of Eqs.\,(\ref{eq:NLS1}) and (\ref{eq:NLS2}) with non-zero $\beta_2$.

The insensitivity to $\beta_2$ is related to an overall pulse width insensitivity of the normalized transmission. Qualitatively, the increase in pump irradiance at fixed energy due to decrease in pulsewidth is compensated by a decrease in interaction length as the pulses walk through each other. In fact, with $\sigma_2 = 0$, the maximum signal of Eq.\,(\ref{eq:analytical}) is completely independent of pulse widths. Since the linear and nonlinear pulse broadening should be negligible over the $\sim 300\,\mu m$ walkoff length, this independence would be true without ignoring $\mathrm{Re}\{\gamma_{ij}\}$ and $\beta_2^{(l)}$. In our case, altering the pulsewidth only causes a slight signal reduction due to a delay shift in the curve moving the peak back to a point where more pump losses have occurred. This only gives an error of around $8\%$ when $\tau_x$ is underestimated by a factor of two. We use this insensitivity to justify ignoring nonlinear refraction in Eqs.\,(\ref{eq:NLS1}) and (\ref{eq:NLS2}).

\subsection*{Data fitting procedure}\label{subsec:fitting}
For TM-TM (TE-TE) sum wavelengths of $743 \,\mathrm{nm}$, $762 \,\mathrm{nm}$, $772 \,\mathrm{nm}$, $784 \,\mathrm{nm}$ ($743 \,\mathrm{nm}$, $762 \,\mathrm{nm}$, $780 \,\mathrm{nm}$) the transmission curves were fit with $\mathrm{Im}\{\gamma_{12}\}$, $\sigma_2$, $\rho$ and $\tau_1$ as free parameters. Averaging the resulting loss coefficents gives $\sigma_{2,TM} = 0.46\,\mathrm{mm^{-1}}$ and $\sigma_{2,TE} = 0.56 \,\mathrm{mm^{-1}}$. A wavelength shift of $25 \,\mathrm{nm}$ was applied to the dispersion curve so the simulated values of $\rho$ more closely match the fits. This adjusment is needed most likely due to the inaccuracy of the spatially averaged approximation to the quantum well index (Eq.\,(\ref{eq:AveragedIndex})). The values presented here were held fixed in fitting the rest of the data. 

The rest of the data points in Figs. \ref{fig:2PATMCurves} and \ref{fig:2PATECurves} were found by measuring the rising edge and peak of the normalized transmission. The pump pulsewidth was taken to be $9\%$ greater than the bandwidth limit (see Sec. \ref{sec:Experiment}), and $\gamma_{12}$ and $\tau_1$ were free fitting parameters. Once again, the effect of this imperfect knowledge of pulsewidths is mitigated by the signal’s insensitivity to pulse width. The $\sigma_2$ and GVM values were held fixed according to the previous fitting procedure. 


\begin{thebibliography}{55}%
\makeatletter
\providecommand \@ifxundefined [1]{%
 \@ifx{#1\undefined}
}%
\providecommand \@ifnum [1]{%
 \ifnum #1\expandafter \@firstoftwo
 \else \expandafter \@secondoftwo
 \fi
}%
\providecommand \@ifx [1]{%
 \ifx #1\expandafter \@firstoftwo
 \else \expandafter \@secondoftwo
 \fi
}%
\providecommand \natexlab [1]{#1}%
\providecommand \enquote  [1]{``#1''}%
\providecommand \bibnamefont  [1]{#1}%
\providecommand \bibfnamefont [1]{#1}%
\providecommand \citenamefont [1]{#1}%
\providecommand \href@noop [0]{\@secondoftwo}%
\providecommand \href [0]{\begingroup \@sanitize@url \@href}%
\providecommand \@href[1]{\@@startlink{#1}\@@href}%
\providecommand \@@href[1]{\endgroup#1\@@endlink}%
\providecommand \@sanitize@url [0]{\catcode `\\12\catcode `\$12\catcode
  `\&12\catcode `\#12\catcode `\^12\catcode `\_12\catcode `\%12\relax}%
\providecommand \@@startlink[1]{}%
\providecommand \@@endlink[0]{}%
\providecommand \url  [0]{\begingroup\@sanitize@url \@url }%
\providecommand \@url [1]{\endgroup\@href {#1}{\urlprefix }}%
\providecommand \urlprefix  [0]{URL }%
\providecommand \Eprint [0]{\href }%
\providecommand \doibase [0]{http://dx.doi.org/}%
\providecommand \selectlanguage [0]{\@gobble}%
\providecommand \bibinfo  [0]{\@secondoftwo}%
\providecommand \bibfield  [0]{\@secondoftwo}%
\providecommand \translation [1]{[#1]}%
\providecommand \BibitemOpen [0]{}%
\providecommand \bibitemStop [0]{}%
\providecommand \bibitemNoStop [0]{.\EOS\space}%
\providecommand \EOS [0]{\spacefactor3000\relax}%
\providecommand \BibitemShut  [1]{\csname bibitem#1\endcsname}%
\let\auto@bib@innerbib\@empty
\bibitem [{\citenamefont {Fishman}\ \emph {et~al.}(2011)\citenamefont
  {Fishman}, \citenamefont {Cirloganu}, \citenamefont {Webster}, \citenamefont
  {Padilha}, \citenamefont {Monroe}, \citenamefont {Hagan},\ and\ \citenamefont
  {{Van Stryland}}}]{Fishman2011}%
  \BibitemOpen
  \bibfield  {author} {\bibinfo {author} {\bibfnamefont {D.~A.}\ \bibnamefont
  {Fishman}}, \bibinfo {author} {\bibfnamefont {C.~M.}\ \bibnamefont
  {Cirloganu}}, \bibinfo {author} {\bibfnamefont {S.}~\bibnamefont {Webster}},
  \bibinfo {author} {\bibfnamefont {L.~A.}\ \bibnamefont {Padilha}}, \bibinfo
  {author} {\bibfnamefont {M.}~\bibnamefont {Monroe}}, \bibinfo {author}
  {\bibfnamefont {D.~J.}\ \bibnamefont {Hagan}}, \ and\ \bibinfo {author}
  {\bibfnamefont {E.~W.}\ \bibnamefont {{Van Stryland}}},\ }\href {\doibase
  10.1038/nphoton.2011.168} {\bibfield  {journal} {\bibinfo  {journal} {Nature
  Photonics}\ }\textbf {\bibinfo {volume} {5}},\ \bibinfo {pages} {561}
  (\bibinfo {year} {2011})}\BibitemShut {NoStop}%
\bibitem [{\citenamefont {Pattanaik}\ \emph
  {et~al.}(2016{\natexlab{a}})\citenamefont {Pattanaik}, \citenamefont
  {Reichert}, \citenamefont {Hagan},\ and\ \citenamefont {{Van
  Stryland}}}]{Pattanaik2016a}%
  \BibitemOpen
  \bibfield  {author} {\bibinfo {author} {\bibfnamefont {H.~S.}\ \bibnamefont
  {Pattanaik}}, \bibinfo {author} {\bibfnamefont {M.}~\bibnamefont {Reichert}},
  \bibinfo {author} {\bibfnamefont {D.~J.}\ \bibnamefont {Hagan}}, \ and\
  \bibinfo {author} {\bibfnamefont {E.~W.}\ \bibnamefont {{Van Stryland}}},\
  }\href {\doibase 10.1364/oe.24.001196} {\bibfield  {journal} {\bibinfo
  {journal} {Optics Express}\ }\textbf {\bibinfo {volume} {24}},\ \bibinfo
  {pages} {1196} (\bibinfo {year} {2016}{\natexlab{a}})}\BibitemShut {NoStop}%
\bibitem [{\citenamefont {Liang}\ \emph {et~al.}(2005)\citenamefont {Liang},
  \citenamefont {Nunes}, \citenamefont {Sakamoto}, \citenamefont {Sasagawa},
  \citenamefont {Kawanishi}, \citenamefont {Tsuchiya}, \citenamefont {Priem},
  \citenamefont {{Van Thourhout}}, \citenamefont {Dumon}, \citenamefont
  {Baets},\ and\ \citenamefont {Tsang}}]{Liang2005}%
  \BibitemOpen
  \bibfield  {author} {\bibinfo {author} {\bibfnamefont {T.~K.}\ \bibnamefont
  {Liang}}, \bibinfo {author} {\bibfnamefont {L.~R.}\ \bibnamefont {Nunes}},
  \bibinfo {author} {\bibfnamefont {T.}~\bibnamefont {Sakamoto}}, \bibinfo
  {author} {\bibfnamefont {K.}~\bibnamefont {Sasagawa}}, \bibinfo {author}
  {\bibfnamefont {T.}~\bibnamefont {Kawanishi}}, \bibinfo {author}
  {\bibfnamefont {M.}~\bibnamefont {Tsuchiya}}, \bibinfo {author}
  {\bibfnamefont {G.~R.~A.}\ \bibnamefont {Priem}}, \bibinfo {author}
  {\bibfnamefont {D.}~\bibnamefont {{Van Thourhout}}}, \bibinfo {author}
  {\bibfnamefont {P.}~\bibnamefont {Dumon}}, \bibinfo {author} {\bibfnamefont
  {R.}~\bibnamefont {Baets}}, \ and\ \bibinfo {author} {\bibfnamefont {H.~K.}\
  \bibnamefont {Tsang}},\ }\href {\doibase 10.1364/opex.13.007298} {\bibfield
  {journal} {\bibinfo  {journal} {Optics Express}\ }\textbf {\bibinfo {volume}
  {13}},\ \bibinfo {pages} {7298} (\bibinfo {year} {2005})}\BibitemShut
  {NoStop}%
\bibitem [{\citenamefont {Reichert}\ \emph {et~al.}(2016)\citenamefont
  {Reichert}, \citenamefont {Smirl}, \citenamefont {Salamo}, \citenamefont
  {Hagan},\ and\ \citenamefont {{Van Stryland}}}]{Reichert2016a}%
  \BibitemOpen
  \bibfield  {author} {\bibinfo {author} {\bibfnamefont {M.}~\bibnamefont
  {Reichert}}, \bibinfo {author} {\bibfnamefont {A.~L.}\ \bibnamefont {Smirl}},
  \bibinfo {author} {\bibfnamefont {G.}~\bibnamefont {Salamo}}, \bibinfo
  {author} {\bibfnamefont {D.~J.}\ \bibnamefont {Hagan}}, \ and\ \bibinfo
  {author} {\bibfnamefont {E.~W.}\ \bibnamefont {{Van Stryland}}},\ }\href
  {\doibase 10.1103/PhysRevLett.117.073602} {\bibfield  {journal} {\bibinfo
  {journal} {Physical Review Letters}\ }\textbf {\bibinfo {volume} {117}},\
  \bibinfo {pages} {073602} (\bibinfo {year} {2016})}\BibitemShut {NoStop}%
\bibitem [{\citenamefont {Melzer}\ \emph {et~al.}(2018)\citenamefont {Melzer},
  \citenamefont {Ruppert}, \citenamefont {Bristow},\ and\ \citenamefont
  {Betz}}]{Melzer2018a}%
  \BibitemOpen
  \bibfield  {author} {\bibinfo {author} {\bibfnamefont {S.}~\bibnamefont
  {Melzer}}, \bibinfo {author} {\bibfnamefont {C.}~\bibnamefont {Ruppert}},
  \bibinfo {author} {\bibfnamefont {A.~D.}\ \bibnamefont {Bristow}}, \ and\
  \bibinfo {author} {\bibfnamefont {M.}~\bibnamefont {Betz}},\ }\href {\doibase
  10.1364/ol.43.005066} {\bibfield  {journal} {\bibinfo  {journal} {Optics
  Letters}\ }\textbf {\bibinfo {volume} {43}},\ \bibinfo {pages} {5066}
  (\bibinfo {year} {2018})}\BibitemShut {NoStop}%
\bibitem [{\citenamefont {Hayat}\ \emph {et~al.}(2008)\citenamefont {Hayat},
  \citenamefont {Ginzburg},\ and\ \citenamefont {Orenstein}}]{Hayat2008}%
  \BibitemOpen
  \bibfield  {author} {\bibinfo {author} {\bibfnamefont {A.}~\bibnamefont
  {Hayat}}, \bibinfo {author} {\bibfnamefont {P.}~\bibnamefont {Ginzburg}}, \
  and\ \bibinfo {author} {\bibfnamefont {M.}~\bibnamefont {Orenstein}},\ }\href
  {\doibase 10.1038/nphoton.2008.28} {\bibfield  {journal} {\bibinfo  {journal}
  {Nature Photonics}\ }\textbf {\bibinfo {volume} {2}},\ \bibinfo {pages} {238}
  (\bibinfo {year} {2008})}\BibitemShut {NoStop}%
\bibitem [{\citenamefont {Ironside}(1992)}]{Ironside1992}%
  \BibitemOpen
  \bibfield  {author} {\bibinfo {author} {\bibfnamefont {C.~N.}\ \bibnamefont
  {Ironside}},\ }\href {https://ieeexplore.ieee.org/document/135201} {\bibfield
   {journal} {\bibinfo  {journal} {IEEE Journal of Quantum Electronics}\
  }\textbf {\bibinfo {volume} {28}},\ \bibinfo {pages} {842} (\bibinfo {year}
  {1992})}\BibitemShut {NoStop}%
\bibitem [{\citenamefont {Gauthier}\ \emph {et~al.}(1992)\citenamefont
  {Gauthier}, \citenamefont {Wu}, \citenamefont {Morin},\ and\ \citenamefont
  {Mossberg}}]{Gauthier1992}%
  \BibitemOpen
  \bibfield  {author} {\bibinfo {author} {\bibfnamefont {D.~J.}\ \bibnamefont
  {Gauthier}}, \bibinfo {author} {\bibfnamefont {Q.}~\bibnamefont {Wu}},
  \bibinfo {author} {\bibfnamefont {S.~E.}\ \bibnamefont {Morin}}, \ and\
  \bibinfo {author} {\bibfnamefont {T.~W.}\ \bibnamefont {Mossberg}},\ }\href
  {\doibase 10.1103/PhysRevLett.68.464} {\bibfield  {journal} {\bibinfo
  {journal} {Physical Review Letters}\ }\textbf {\bibinfo {volume} {68}},\
  \bibinfo {pages} {464} (\bibinfo {year} {1992})}\BibitemShut {NoStop}%
\bibitem [{\citenamefont {Hayat}\ \emph {et~al.}(2011)\citenamefont {Hayat},
  \citenamefont {Nevet}, \citenamefont {Ginzburg},\ and\ \citenamefont
  {Orenstein}}]{Hayat2011}%
  \BibitemOpen
  \bibfield  {author} {\bibinfo {author} {\bibfnamefont {A.}~\bibnamefont
  {Hayat}}, \bibinfo {author} {\bibfnamefont {A.}~\bibnamefont {Nevet}},
  \bibinfo {author} {\bibfnamefont {P.}~\bibnamefont {Ginzburg}}, \ and\
  \bibinfo {author} {\bibfnamefont {M.}~\bibnamefont {Orenstein}},\ }\href
  {https://iopscience.iop.org/article/10.1088/0268-1242/26/8/083001} {\bibfield
   {journal} {\bibinfo  {journal} {Semiconductor Science and Technology}\
  }\textbf {\bibinfo {volume} {26}} (\bibinfo {year} {2011})}\BibitemShut
  {NoStop}%
\bibitem [{\citenamefont {Poulvellarie}\ \emph {et~al.}(2018)\citenamefont
  {Poulvellarie}, \citenamefont {Ciret}, \citenamefont {Kuyken}, \citenamefont
  {Leo},\ and\ \citenamefont {Gorza}}]{Poulvellarie2018}%
  \BibitemOpen
  \bibfield  {author} {\bibinfo {author} {\bibfnamefont {N.}~\bibnamefont
  {Poulvellarie}}, \bibinfo {author} {\bibfnamefont {C.}~\bibnamefont {Ciret}},
  \bibinfo {author} {\bibfnamefont {B.}~\bibnamefont {Kuyken}}, \bibinfo
  {author} {\bibfnamefont {F.}~\bibnamefont {Leo}}, \ and\ \bibinfo {author}
  {\bibfnamefont {S.~P.}\ \bibnamefont {Gorza}},\ }\href {\doibase
  10.1103/PhysRevApplied.10.024033} {\bibfield  {journal} {\bibinfo  {journal}
  {Physical Review Applied}\ }\textbf {\bibinfo {volume} {10}},\ \bibinfo
  {pages} {24033} (\bibinfo {year} {2018})}\BibitemShut {NoStop}%
\bibitem [{\citenamefont {Spector}(1987)}]{Spector1987b}%
  \BibitemOpen
  \bibfield  {author} {\bibinfo {author} {\bibfnamefont {H.~N.}\ \bibnamefont
  {Spector}},\ }\href {\doibase 10.1103/PhysRevB.35.5876} {\bibfield  {journal}
  {\bibinfo  {journal} {Physical Review B}\ }\textbf {\bibinfo {volume} {35}},\
  \bibinfo {pages} {5876} (\bibinfo {year} {1987})}\BibitemShut {NoStop}%
\bibitem [{\citenamefont {Pasquarello}\ and\ \citenamefont
  {Quattropani}(1988)}]{Pasquarello1988}%
  \BibitemOpen
  \bibfield  {author} {\bibinfo {author} {\bibfnamefont {A.}~\bibnamefont
  {Pasquarello}}\ and\ \bibinfo {author} {\bibfnamefont {A.}~\bibnamefont
  {Quattropani}},\ }\href {\doibase 10.1103/PhysRevB.38.6206} {\bibfield
  {journal} {\bibinfo  {journal} {Physical Review B}\ }\textbf {\bibinfo
  {volume} {38}},\ \bibinfo {pages} {6206} (\bibinfo {year}
  {1988})}\BibitemShut {NoStop}%
\bibitem [{\citenamefont {Nithisoontorn}\ \emph {et~al.}(1989)\citenamefont
  {Nithisoontorn}, \citenamefont {Unterrainer}, \citenamefont {Michaelis},
  \citenamefont {Sawaki}, \citenamefont {Gornik},\ and\ \citenamefont
  {Kano}}]{Nithisoontorn1989}%
  \BibitemOpen
  \bibfield  {author} {\bibinfo {author} {\bibfnamefont {M.}~\bibnamefont
  {Nithisoontorn}}, \bibinfo {author} {\bibfnamefont {K.}~\bibnamefont
  {Unterrainer}}, \bibinfo {author} {\bibfnamefont {S.}~\bibnamefont
  {Michaelis}}, \bibinfo {author} {\bibfnamefont {N.}~\bibnamefont {Sawaki}},
  \bibinfo {author} {\bibfnamefont {E.}~\bibnamefont {Gornik}}, \ and\ \bibinfo
  {author} {\bibfnamefont {H.}~\bibnamefont {Kano}},\ }\href {\doibase
  10.1103/PhysRevLett.62.3078} {\bibfield  {journal} {\bibinfo  {journal}
  {Physical Review Letters}\ }\textbf {\bibinfo {volume} {62}},\ \bibinfo
  {pages} {3078} (\bibinfo {year} {1989})}\BibitemShut {NoStop}%
\bibitem [{\citenamefont {Shimizu}(1989)}]{Shimizu1989}%
  \BibitemOpen
  \bibfield  {author} {\bibinfo {author} {\bibfnamefont {A.}~\bibnamefont
  {Shimizu}},\ }\href {\doibase 10.1103/PhysRevB.40.1403} {\bibfield  {journal}
  {\bibinfo  {journal} {Physical Review B}\ }\textbf {\bibinfo {volume} {40}},\
  \bibinfo {pages} {1403} (\bibinfo {year} {1989})}\BibitemShut {NoStop}%
\bibitem [{\citenamefont {Tai}\ \emph {et~al.}(1989)\citenamefont {Tai},
  \citenamefont {Mysyrowicz}, \citenamefont {Fischer}, \citenamefont
  {Slusher},\ and\ \citenamefont {Cho}}]{Tai1989}%
  \BibitemOpen
  \bibfield  {author} {\bibinfo {author} {\bibfnamefont {K.}~\bibnamefont
  {Tai}}, \bibinfo {author} {\bibfnamefont {A.}~\bibnamefont {Mysyrowicz}},
  \bibinfo {author} {\bibfnamefont {R.~J.}\ \bibnamefont {Fischer}}, \bibinfo
  {author} {\bibfnamefont {R.~E.}\ \bibnamefont {Slusher}}, \ and\ \bibinfo
  {author} {\bibfnamefont {A.~Y.}\ \bibnamefont {Cho}},\ }\href {\doibase
  10.1103/PhysRevLett.62.1784} {\bibfield  {journal} {\bibinfo  {journal}
  {Physical Review Letters}\ }\textbf {\bibinfo {volume} {62}},\ \bibinfo
  {pages} {1784} (\bibinfo {year} {1989})}\BibitemShut {NoStop}%
\bibitem [{\citenamefont {Yang}\ \emph {et~al.}(1993)\citenamefont {Yang},
  \citenamefont {Villeneuve}, \citenamefont {Stegeman}, \citenamefont {Lin},\
  and\ \citenamefont {Lin}}]{Yang1993}%
  \BibitemOpen
  \bibfield  {author} {\bibinfo {author} {\bibfnamefont {C.~C.}\ \bibnamefont
  {Yang}}, \bibinfo {author} {\bibfnamefont {A.}~\bibnamefont {Villeneuve}},
  \bibinfo {author} {\bibfnamefont {G.~I.}\ \bibnamefont {Stegeman}}, \bibinfo
  {author} {\bibfnamefont {C.~H.}\ \bibnamefont {Lin}}, \ and\ \bibinfo
  {author} {\bibfnamefont {H.~H.}\ \bibnamefont {Lin}},\ }\href {\doibase
  10.1109/3.259409} {\bibfield  {journal} {\bibinfo  {journal} {IEEE Journal of
  Quantum Electronics}\ }\textbf {\bibinfo {volume} {29}},\ \bibinfo {pages}
  {2934} (\bibinfo {year} {1993})}\BibitemShut {NoStop}%
\bibitem [{\citenamefont {Pasquarello}\ and\ \citenamefont
  {Quattropani}(1990)}]{Pasquarello1990a}%
  \BibitemOpen
  \bibfield  {author} {\bibinfo {author} {\bibfnamefont {A.}~\bibnamefont
  {Pasquarello}}\ and\ \bibinfo {author} {\bibfnamefont {A.}~\bibnamefont
  {Quattropani}},\ }\href {\doibase 10.1103/PhysRevB.42.9073} {\bibfield
  {journal} {\bibinfo  {journal} {Physical Review B}\ }\textbf {\bibinfo
  {volume} {42}},\ \bibinfo {pages} {9073} (\bibinfo {year}
  {1990})}\BibitemShut {NoStop}%
\bibitem [{\citenamefont {Pattanaik}\ \emph
  {et~al.}(2016{\natexlab{b}})\citenamefont {Pattanaik}, \citenamefont
  {Reichert}, \citenamefont {Khurgin}, \citenamefont {Hagan},\ and\
  \citenamefont {{Van Stryland}}}]{Pattanaik2016}%
  \BibitemOpen
  \bibfield  {author} {\bibinfo {author} {\bibfnamefont {H.~S.}\ \bibnamefont
  {Pattanaik}}, \bibinfo {author} {\bibfnamefont {M.}~\bibnamefont {Reichert}},
  \bibinfo {author} {\bibfnamefont {J.~B.}\ \bibnamefont {Khurgin}}, \bibinfo
  {author} {\bibfnamefont {D.~J.}\ \bibnamefont {Hagan}}, \ and\ \bibinfo
  {author} {\bibfnamefont {E.~W.}\ \bibnamefont {{Van Stryland}}},\ }\href
  {\doibase 10.1109/JQE.2016.2526611} {\bibfield  {journal} {\bibinfo
  {journal} {IEEE Journal of Quantum Electronics}\ }\textbf {\bibinfo {volume}
  {52}},\ \bibinfo {pages} {1} (\bibinfo {year}
  {2016}{\natexlab{b}})}\BibitemShut {NoStop}%
\bibitem [{\citenamefont {Khurgin}(1994)}]{Khurgin1994}%
  \BibitemOpen
  \bibfield  {author} {\bibinfo {author} {\bibfnamefont {J.~B.}\ \bibnamefont
  {Khurgin}},\ }\href {\doibase 10.1364/josab.11.000624} {\bibfield  {journal}
  {\bibinfo  {journal} {Journal of the Optical Society of America B}\ }\textbf
  {\bibinfo {volume} {11}},\ \bibinfo {pages} {624} (\bibinfo {year}
  {1994})}\BibitemShut {NoStop}%
\bibitem [{Sup()}]{Supp}%
  \BibitemOpen
  \href@noop {} {}\bibinfo {note} {See Supplemental Material at [URL will be
  inserted by publisher] for derivations of the ND-2PA coefficient formula,
  energy-dependent effective masses, and the analytical expression for
  nonlinear propagation.}\BibitemShut {Stop}%
\bibitem [{\citenamefont {Chuang}(1995)}]{Chuang1995}%
  \BibitemOpen
  \bibfield  {author} {\bibinfo {author} {\bibfnamefont {S.~L.}\ \bibnamefont
  {Chuang}},\ }\href {\doibase 10.1063/1.2807693} {\emph {\bibinfo {title}
  {{Physics of Optoelectronic Devices}}}}\ (\bibinfo  {publisher} {Wiley},\
  \bibinfo {year} {1995})\BibitemShut {NoStop}%
\bibitem [{\citenamefont {Lee}\ and\ \citenamefont {Fan}(1974)}]{Lee1974}%
  \BibitemOpen
  \bibfield  {author} {\bibinfo {author} {\bibfnamefont {C.~C.}\ \bibnamefont
  {Lee}}\ and\ \bibinfo {author} {\bibfnamefont {H.~Y.}\ \bibnamefont {Fan}},\
  }\href {\doibase 10.1103/PhysRevB.9.3502} {\bibfield  {journal} {\bibinfo
  {journal} {Physical Review B}\ }\textbf {\bibinfo {volume} {9}},\ \bibinfo
  {pages} {3502} (\bibinfo {year} {1974})}\BibitemShut {NoStop}%
\bibitem [{\citenamefont {Basov}\ \emph {et~al.}(1966)\citenamefont {Basov},
  \citenamefont {Grasyuk}, \citenamefont {Zubarev}, \citenamefont {Katulin},\
  and\ \citenamefont {Krokhin}}]{Basov1966}%
  \BibitemOpen
  \bibfield  {author} {\bibinfo {author} {\bibfnamefont {N.~G.}\ \bibnamefont
  {Basov}}, \bibinfo {author} {\bibfnamefont {A.~Z.}\ \bibnamefont {Grasyuk}},
  \bibinfo {author} {\bibfnamefont {I.~G.}\ \bibnamefont {Zubarev}}, \bibinfo
  {author} {\bibfnamefont {V.~A.}\ \bibnamefont {Katulin}}, \ and\ \bibinfo
  {author} {\bibfnamefont {.~N.}\ \bibnamefont {Krokhin}},\ }\href
  {http://www.jetp.ac.ru/cgi-bin/e/index/e/23/3/p366?a=list} {\bibfield
  {journal} {\bibinfo  {journal} {J. Exptl. Theoret. Phys. (U.S.S.R.)}\
  }\textbf {\bibinfo {volume} {23}},\ \bibinfo {pages} {551} (\bibinfo {year}
  {1966})}\BibitemShut {NoStop}%
\bibitem [{\citenamefont {Sheik-Bahae}\ and\ \citenamefont {{Van
  Stryland}}(1999)}]{Sheik-Bahae1998}%
  \BibitemOpen
  \bibfield  {author} {\bibinfo {author} {\bibfnamefont {M.}~\bibnamefont
  {Sheik-Bahae}}\ and\ \bibinfo {author} {\bibfnamefont {E.~W.}\ \bibnamefont
  {{Van Stryland}}},\ }in\ \href {\doibase 10.1016/S0080-8784(08)62723-4}
  {\emph {\bibinfo {booktitle} {Nonlinear Optics in Semiconductors I}}}\
  (\bibinfo  {publisher} {Elsevier Masson SAS},\ \bibinfo {year} {1999})\
  Chap.~\bibinfo {chapter} {4}, pp.\ \bibinfo {pages} {257--318}\BibitemShut
  {NoStop}%
\bibitem [{\citenamefont {Hutchings}\ and\ \citenamefont {{Van
  Stryland}}(1992)}]{Hutchings1992c}%
  \BibitemOpen
  \bibfield  {author} {\bibinfo {author} {\bibfnamefont {D.~C.}\ \bibnamefont
  {Hutchings}}\ and\ \bibinfo {author} {\bibfnamefont {E.~W.}\ \bibnamefont
  {{Van Stryland}}},\ }\href {\doibase 10.1364/josab.9.002065} {\bibfield
  {journal} {\bibinfo  {journal} {Journal of the Optical Society of America B}\
  }\textbf {\bibinfo {volume} {9}},\ \bibinfo {pages} {2065} (\bibinfo {year}
  {1992})}\BibitemShut {NoStop}%
\bibitem [{\citenamefont {Luttinger}\ and\ \citenamefont
  {Kohn}(1955)}]{Kohn1954}%
  \BibitemOpen
  \bibfield  {author} {\bibinfo {author} {\bibfnamefont {J.~M.}\ \bibnamefont
  {Luttinger}}\ and\ \bibinfo {author} {\bibfnamefont {W.}~\bibnamefont
  {Kohn}},\ }\href {\doibase 10.1103/PhysRev.97.869} {\bibfield  {journal}
  {\bibinfo  {journal} {Physical Review}\ }\textbf {\bibinfo {volume} {97}},\
  \bibinfo {pages} {869} (\bibinfo {year} {1955})}\BibitemShut {NoStop}%
\bibitem [{\citenamefont {Bastard}(1981)}]{Bastard1981}%
  \BibitemOpen
  \bibfield  {author} {\bibinfo {author} {\bibfnamefont {G.}~\bibnamefont
  {Bastard}},\ }\href {\doibase 10.1103/PhysRevB.24.5693} {\bibfield  {journal}
  {\bibinfo  {journal} {Physical Review B}\ }\textbf {\bibinfo {volume} {24}},\
  \bibinfo {pages} {5693} (\bibinfo {year} {1981})}\BibitemShut {NoStop}%
\bibitem [{\citenamefont {Bastard}(1976)}]{Bastard1976}%
  \BibitemOpen
  \bibfield  {author} {\bibinfo {author} {\bibfnamefont {G.}~\bibnamefont
  {Bastard}},\ }\href@noop {} {\emph {\bibinfo {title} {{Wave Mechanics Applied
  to Semiconductor Heterostructures}}}}\ (\bibinfo  {publisher}
  {Wiley-Interscience},\ \bibinfo {year} {1976})\BibitemShut {NoStop}%
\bibitem [{\citenamefont {L{\"{o}}wdin}(1951)}]{Lowdin1951}%
  \BibitemOpen
  \bibfield  {author} {\bibinfo {author} {\bibfnamefont {P.~O.}\ \bibnamefont
  {L{\"{o}}wdin}},\ }\href {\doibase 10.1063/1.1748067} {\bibfield  {journal}
  {\bibinfo  {journal} {The Journal of Chemical Physics}\ }\textbf {\bibinfo
  {volume} {19}},\ \bibinfo {pages} {1396} (\bibinfo {year}
  {1951})}\BibitemShut {NoStop}%
\bibitem [{\citenamefont {Meney}\ \emph {et~al.}(1994)\citenamefont {Meney},
  \citenamefont {Gonul},\ and\ \citenamefont {O'Reilly}}]{Meney1994}%
  \BibitemOpen
  \bibfield  {author} {\bibinfo {author} {\bibfnamefont {A.~T.}\ \bibnamefont
  {Meney}}, \bibinfo {author} {\bibfnamefont {B.}~\bibnamefont {Gonul}}, \ and\
  \bibinfo {author} {\bibfnamefont {E.}~\bibnamefont {O'Reilly}},\ }\href
  {\doibase https://doi.org/10.1103/PhysRevB.50.10893} {\bibfield  {journal}
  {\bibinfo  {journal} {Physical Review B}\ }\textbf {\bibinfo {volume} {50}},\
  \bibinfo {pages} {893} (\bibinfo {year} {1994})}\BibitemShut {NoStop}%
\bibitem [{\citenamefont {Wherrett}(1984)}]{Wherrett1984a}%
  \BibitemOpen
  \bibfield  {author} {\bibinfo {author} {\bibfnamefont {B.~S.}\ \bibnamefont
  {Wherrett}},\ }\href {\doibase 10.1364/josab.1.000067} {\bibfield  {journal}
  {\bibinfo  {journal} {Journal of the Optical Society of America B}\ }\textbf
  {\bibinfo {volume} {1}},\ \bibinfo {pages} {67} (\bibinfo {year}
  {1984})}\BibitemShut {NoStop}%
\bibitem [{\citenamefont {Faist}(2013)}]{Faist2013}%
  \BibitemOpen
  \bibfield  {author} {\bibinfo {author} {\bibfnamefont {J.}~\bibnamefont
  {Faist}},\ }\href
  {https://global.oup.com/academic/product/quantum-cascade-lasers-9780198795889?cc=us{\&}lang=en{\&}}
  {\emph {\bibinfo {title} {{Quantum cascade lasers}}}}\ (\bibinfo  {publisher}
  {Oxford University Press},\ \bibinfo {year} {2013})\BibitemShut {NoStop}%
\bibitem [{\citenamefont {Ashcroft}\ and\ \citenamefont
  {Mermin}(1976)}]{NeilW.Ashcroft1976}%
  \BibitemOpen
  \bibfield  {author} {\bibinfo {author} {\bibfnamefont {N.~W.}\ \bibnamefont
  {Ashcroft}}\ and\ \bibinfo {author} {\bibfnamefont {N.~D.}\ \bibnamefont
  {Mermin}},\ }\href {https://www.cengage.co.uk/books/9780030839931/} {\emph
  {\bibinfo {title} {{Solid State Physics}}}}\ (\bibinfo  {publisher}
  {Cengage},\ \bibinfo {year} {1976})\BibitemShut {NoStop}%
\bibitem [{\citenamefont {Yamanishi}\ and\ \citenamefont
  {Suemune}(1984)}]{Yamanishi1984}%
  \BibitemOpen
  \bibfield  {author} {\bibinfo {author} {\bibfnamefont {M.}~\bibnamefont
  {Yamanishi}}\ and\ \bibinfo {author} {\bibfnamefont {I.}~\bibnamefont
  {Suemune}},\ }\href {\doibase 10.1143/JJAP.23.L35} {\bibfield  {journal}
  {\bibinfo  {journal} {Japanese Journal of Applied Physics}\ }\textbf
  {\bibinfo {volume} {23}},\ \bibinfo {pages} {35} (\bibinfo {year}
  {1984})}\BibitemShut {NoStop}%
\bibitem [{\citenamefont {Hermann}\ and\ \citenamefont
  {Weisbuch}(1977)}]{Hermann1977}%
  \BibitemOpen
  \bibfield  {author} {\bibinfo {author} {\bibfnamefont {C.}~\bibnamefont
  {Hermann}}\ and\ \bibinfo {author} {\bibfnamefont {C.}~\bibnamefont
  {Weisbuch}},\ }\href {\doibase 10.1103/PhysRevB.15.823} {\bibfield  {journal}
  {\bibinfo  {journal} {Physical Review B}\ }\textbf {\bibinfo {volume} {15}},\
  \bibinfo {pages} {823} (\bibinfo {year} {1977})}\BibitemShut {NoStop}%
\bibitem [{\citenamefont {Chuang}(1991)}]{Chuang1991}%
  \BibitemOpen
  \bibfield  {author} {\bibinfo {author} {\bibfnamefont {S.~L.}\ \bibnamefont
  {Chuang}},\ }\href {\doibase 10.1103/PhysRevB.43.9649} {\bibfield  {journal}
  {\bibinfo  {journal} {Physical Review B}\ }\textbf {\bibinfo {volume} {43}},\
  \bibinfo {pages} {9649} (\bibinfo {year} {1991})}\BibitemShut {NoStop}%
\bibitem [{\citenamefont {H{\"{u}}bner}\ \emph {et~al.}(2009)\citenamefont
  {H{\"{u}}bner}, \citenamefont {D{\"{o}}hrmann}, \citenamefont
  {H{\"{a}}gele},\ and\ \citenamefont {Oestreich}}]{Hubner2009}%
  \BibitemOpen
  \bibfield  {author} {\bibinfo {author} {\bibfnamefont {J.}~\bibnamefont
  {H{\"{u}}bner}}, \bibinfo {author} {\bibfnamefont {S.}~\bibnamefont
  {D{\"{o}}hrmann}}, \bibinfo {author} {\bibfnamefont {D.}~\bibnamefont
  {H{\"{a}}gele}}, \ and\ \bibinfo {author} {\bibfnamefont {M.}~\bibnamefont
  {Oestreich}},\ }\href {\doibase 10.1103/PhysRevB.79.193307} {\bibfield
  {journal} {\bibinfo  {journal} {Physical Review B - Condensed Matter and
  Materials Physics}\ }\textbf {\bibinfo {volume} {79}},\ \bibinfo {pages} {1}
  (\bibinfo {year} {2009})}\BibitemShut {NoStop}%
\bibitem [{\citenamefont {Cirloganu}\ \emph {et~al.}(2011)\citenamefont
  {Cirloganu}, \citenamefont {Padilha}, \citenamefont {Fishman}, \citenamefont
  {Webster}, \citenamefont {Hagan},\ and\ \citenamefont {{Van
  Stryland}}}]{Cirloganu2011a}%
  \BibitemOpen
  \bibfield  {author} {\bibinfo {author} {\bibfnamefont {C.~M.}\ \bibnamefont
  {Cirloganu}}, \bibinfo {author} {\bibfnamefont {L.~A.}\ \bibnamefont
  {Padilha}}, \bibinfo {author} {\bibfnamefont {D.~A.}\ \bibnamefont
  {Fishman}}, \bibinfo {author} {\bibfnamefont {S.}~\bibnamefont {Webster}},
  \bibinfo {author} {\bibfnamefont {D.~J.}\ \bibnamefont {Hagan}}, \ and\
  \bibinfo {author} {\bibfnamefont {E.~W.}\ \bibnamefont {{Van Stryland}}},\
  }\href {\doibase 10.1364/oe.19.022951} {\bibfield  {journal} {\bibinfo
  {journal} {Optics Express}\ }\textbf {\bibinfo {volume} {19}},\ \bibinfo
  {pages} {22951} (\bibinfo {year} {2011})}\BibitemShut {NoStop}%
\bibitem [{\citenamefont {Kane}(1957)}]{Kane1957}%
  \BibitemOpen
  \bibfield  {author} {\bibinfo {author} {\bibfnamefont {E.~O.}\ \bibnamefont
  {Kane}},\ }\href {\doibase 10.1016/0022-3697(57)90013-6} {\bibfield
  {journal} {\bibinfo  {journal} {Journal of Physics and Chemistry of Solids}\
  }\textbf {\bibinfo {volume} {1}},\ \bibinfo {pages} {249} (\bibinfo {year}
  {1957})}\BibitemShut {NoStop}%
\bibitem [{\citenamefont {Yu}\ and\ \citenamefont {Cardona}(2010)}]{Yu1980}%
  \BibitemOpen
  \bibfield  {author} {\bibinfo {author} {\bibfnamefont {P.~Y.}\ \bibnamefont
  {Yu}}\ and\ \bibinfo {author} {\bibfnamefont {M.}~\bibnamefont {Cardona}},\
  }\href {\doibase 10.1007/978-3-642-00710-1} {\emph {\bibinfo {title}
  {{Fundamentals of semiconductors}}}}\ (\bibinfo  {publisher} {Springer},\
  \bibinfo {address} {New York},\ \bibinfo {year} {2010})\BibitemShut {NoStop}%
\bibitem [{\citenamefont {Varshni}(1967)}]{Varshni1967}%
  \BibitemOpen
  \bibfield  {author} {\bibinfo {author} {\bibfnamefont {Y.}~\bibnamefont
  {Varshni}},\ }\href {\doibase doi: 10.1016/0031-8914(67)90062-6} {\bibfield
  {journal} {\bibinfo  {journal} {Physica}\ }\textbf {\bibinfo {volume} {34}},\
  \bibinfo {pages} {149} (\bibinfo {year} {1967})}\BibitemShut {NoStop}%
\bibitem [{\citenamefont {{El Allali}}\ \emph {et~al.}(1993)\citenamefont {{El
  Allali}}, \citenamefont {Sorensen}, \citenamefont {Veje},\ and\ \citenamefont
  {Tidemand-Petersson}}]{ElAllali1993}%
  \BibitemOpen
  \bibfield  {author} {\bibinfo {author} {\bibfnamefont {M.}~\bibnamefont {{El
  Allali}}}, \bibinfo {author} {\bibfnamefont {C.~B.}\ \bibnamefont
  {Sorensen}}, \bibinfo {author} {\bibfnamefont {E.}~\bibnamefont {Veje}}, \
  and\ \bibinfo {author} {\bibfnamefont {P.}~\bibnamefont
  {Tidemand-Petersson}},\ }\href {\doibase 10.1103/PhysRevB.48.4398} {\bibfield
   {journal} {\bibinfo  {journal} {Physical Review B}\ }\textbf {\bibinfo
  {volume} {48}},\ \bibinfo {pages} {4398} (\bibinfo {year}
  {1993})}\BibitemShut {NoStop}%
\bibitem [{\citenamefont {Vurgaftman}\ \emph {et~al.}(2001)\citenamefont
  {Vurgaftman}, \citenamefont {Meyer},\ and\ \citenamefont
  {Ram-Mohan}}]{Vurgaftman2001}%
  \BibitemOpen
  \bibfield  {author} {\bibinfo {author} {\bibfnamefont {I.}~\bibnamefont
  {Vurgaftman}}, \bibinfo {author} {\bibfnamefont {J.~R.}\ \bibnamefont
  {Meyer}}, \ and\ \bibinfo {author} {\bibfnamefont {L.~R.}\ \bibnamefont
  {Ram-Mohan}},\ }\href {\doibase 10.1063/1.1368156} {\bibfield  {journal}
  {\bibinfo  {journal} {Journal of Applied Physics}\ }\textbf {\bibinfo
  {volume} {89}},\ \bibinfo {pages} {5815} (\bibinfo {year}
  {2001})}\BibitemShut {NoStop}%
\bibitem [{\citenamefont {Cardona}\ \emph {et~al.}(1967)\citenamefont
  {Cardona}, \citenamefont {Shaklee},\ and\ \citenamefont
  {Pollak}}]{Cardona1967}%
  \BibitemOpen
  \bibfield  {author} {\bibinfo {author} {\bibfnamefont {M.}~\bibnamefont
  {Cardona}}, \bibinfo {author} {\bibfnamefont {K.~L.}\ \bibnamefont
  {Shaklee}}, \ and\ \bibinfo {author} {\bibfnamefont {F.~H.}\ \bibnamefont
  {Pollak}},\ }\href {\doibase 10.1103/PhysRev.154.696} {\bibfield  {journal}
  {\bibinfo  {journal} {Physical Review}\ }\textbf {\bibinfo {volume} {154}},\
  \bibinfo {pages} {696} (\bibinfo {year} {1967})}\BibitemShut {NoStop}%
\bibitem [{\citenamefont {Balslev}(1969)}]{Balslev1969}%
  \BibitemOpen
  \bibfield  {author} {\bibinfo {author} {\bibfnamefont {I.}~\bibnamefont
  {Balslev}},\ }\href {\doibase https://doi.org/10.1103/PhysRev.177.1173}
  {\bibfield  {journal} {\bibinfo  {journal} {Physical Review}\ }\textbf
  {\bibinfo {volume} {177}},\ \bibinfo {pages} {1173} (\bibinfo {year}
  {1969})}\BibitemShut {NoStop}%
\bibitem [{\citenamefont {Lawaetz}(1971)}]{Lawaetz1971}%
  \BibitemOpen
  \bibfield  {author} {\bibinfo {author} {\bibfnamefont {P.}~\bibnamefont
  {Lawaetz}},\ }\href {\doibase 10.1103/PhysRevB.4.3460} {\bibfield  {journal}
  {\bibinfo  {journal} {Physical Review B}\ }\textbf {\bibinfo {volume} {4}},\
  \bibinfo {pages} {3460} (\bibinfo {year} {1971})}\BibitemShut {NoStop}%
\bibitem [{\citenamefont {Pfeffer}\ and\ \citenamefont
  {Zawadzki}(1996)}]{Pfeffer1996}%
  \BibitemOpen
  \bibfield  {author} {\bibinfo {author} {\bibfnamefont {P.}~\bibnamefont
  {Pfeffer}}\ and\ \bibinfo {author} {\bibfnamefont {W.}~\bibnamefont
  {Zawadzki}},\ }\href {\doibase 10.1103/PhysRevB.53.12813} {\bibfield
  {journal} {\bibinfo  {journal} {Physical Review B}\ }\textbf {\bibinfo
  {volume} {53}},\ \bibinfo {pages} {12813} (\bibinfo {year}
  {1996})}\BibitemShut {NoStop}%
\bibitem [{\citenamefont {Eppenga}\ \emph {et~al.}(1987)\citenamefont
  {Eppenga}, \citenamefont {Schuurmans},\ and\ \citenamefont
  {Colak}}]{Eppenga1987}%
  \BibitemOpen
  \bibfield  {author} {\bibinfo {author} {\bibfnamefont {R.}~\bibnamefont
  {Eppenga}}, \bibinfo {author} {\bibfnamefont {M.~H.}\ \bibnamefont
  {Schuurmans}}, \ and\ \bibinfo {author} {\bibfnamefont {S.}~\bibnamefont
  {Colak}},\ }\href
  {https://journals.aps.org/prb/abstract/10.1103/PhysRevB.36.1554} {\bibfield
  {journal} {\bibinfo  {journal} {Physical Review B}\ }\textbf {\bibinfo
  {volume} {36}} (\bibinfo {year} {1987})}\BibitemShut {NoStop}%
\bibitem [{\citenamefont {Vrehen}(1968)}]{Vrehen1968}%
  \BibitemOpen
  \bibfield  {author} {\bibinfo {author} {\bibfnamefont {Q.~H.~F.}\
  \bibnamefont {Vrehen}},\ }\href {\doibase
  https://doi.org/10.1016/0022-3697(68)90263-1} {\bibfield  {journal} {\bibinfo
   {journal} {Journal of Physics and Chemistry of Solids}\ }\textbf {\bibinfo
  {volume} {29}},\ \bibinfo {pages} {129} (\bibinfo {year} {1968})}\BibitemShut
  {NoStop}%
\bibitem [{\citenamefont {Aspnes}\ and\ \citenamefont
  {Studna}(1973)}]{Aspnes1973}%
  \BibitemOpen
  \bibfield  {author} {\bibinfo {author} {\bibfnamefont {D.~E.}\ \bibnamefont
  {Aspnes}}\ and\ \bibinfo {author} {\bibfnamefont {A.~A.}\ \bibnamefont
  {Studna}},\ }\href {\doibase 10.1103/PhysRevB.7.4605} {\bibfield  {journal}
  {\bibinfo  {journal} {Physical Review B}\ }\textbf {\bibinfo {volume} {7}},\
  \bibinfo {pages} {4605} (\bibinfo {year} {1973})}\BibitemShut {NoStop}%
\bibitem [{\citenamefont {Harrison}\ and\ \citenamefont
  {Valavanis}(2005)}]{Harrison2005}%
  \BibitemOpen
  \bibfield  {author} {\bibinfo {author} {\bibfnamefont {P.~P.}\ \bibnamefont
  {Harrison}}\ and\ \bibinfo {author} {\bibfnamefont {A.}~\bibnamefont
  {Valavanis}},\ }\href
  {https://onlinelibrary.wiley.com/doi/book/10.1002/9781118923337} {\emph
  {\bibinfo {title} {{Quantum wells, wires and dots}}}},\ \bibinfo {edition}
  {2nd}\ ed.\ (\bibinfo  {publisher} {Wiley},\ \bibinfo {address} {Sussex,
  England},\ \bibinfo {year} {2005})\BibitemShut {NoStop}%
\bibitem [{\citenamefont {Adachi}(1989)}]{Adachi1989}%
  \BibitemOpen
  \bibfield  {author} {\bibinfo {author} {\bibfnamefont {S.}~\bibnamefont
  {Adachi}},\ }\href {\doibase 10.1063/1.343580} {\bibfield  {journal}
  {\bibinfo  {journal} {Journal of Applied Physics}\ }\textbf {\bibinfo
  {volume} {66}},\ \bibinfo {pages} {6030} (\bibinfo {year}
  {1989})}\BibitemShut {NoStop}%
\bibitem [{Lum()}]{Lumerical}%
  \BibitemOpen
  \href {https://www.lumerical.com/products/} {\enquote {\bibinfo {title}
  {{Lumerical MODE Solutions}},}\ }\BibitemShut {NoStop}%
\bibitem [{\citenamefont {Lin}\ \emph {et~al.}(2007)\citenamefont {Lin},
  \citenamefont {Painter},\ and\ \citenamefont {Agrawal}}]{Lin2007a}%
  \BibitemOpen
  \bibfield  {author} {\bibinfo {author} {\bibfnamefont {Q.}~\bibnamefont
  {Lin}}, \bibinfo {author} {\bibfnamefont {O.~J.}\ \bibnamefont {Painter}}, \
  and\ \bibinfo {author} {\bibfnamefont {G.~P.}\ \bibnamefont {Agrawal}},\
  }\href {\doibase 10.1364/OE.15.016604} {\bibfield  {journal} {\bibinfo
  {journal} {Optics Express}\ }\textbf {\bibinfo {volume} {15}},\ \bibinfo
  {pages} {16604} (\bibinfo {year} {2007})}\BibitemShut {NoStop}%
\bibitem [{\citenamefont {Agrawal}(2012)}]{Agrawal2012}%
  \BibitemOpen
  \bibfield  {author} {\bibinfo {author} {\bibfnamefont {G.~P.}\ \bibnamefont
  {Agrawal}},\ }\href
  {https://www.elsevier.com/books/nonlinear-fiber-optics/agrawal/978-0-12-397023-7}
  {\emph {\bibinfo {title} {{Nonlinear Fiber Optics}}}}\ (\bibinfo  {publisher}
  {Academic Press},\ \bibinfo {year} {2012})\ p.\ \bibinfo {pages}
  {631}\BibitemShut {NoStop}%
\end{thebibliography}
\end{document}


\section{Derivation of 2PA Coefficients}
We start with the second order transition rate per unit volume
    \begin{equation}
        W = \frac{2\pi}{\hbar}\frac{1}{V}\sum_{\mathbf{k}_t}\sum_{cv}\left|\frac{H_{ci}^{2}H_{iv}^{1}}{E_{iv}-\hbar\omega_1}+\frac{H_{ci}^{1}H_{iv}^{2}}{E_{iv}-\hbar\omega_2}\right|^2 \delta(\hbar\omega_1 + \hbar\omega_2 - E_{cv}),
    \end{equation}
where 
\begin{equation}
    H_{mn} = \frac{e}{2m_0}\mathbf{A}_0\cdot \mathbf{p}
\end{equation}
for a harmonic vector potential of magnitude $\mathbf{A}_0$. We can normalize all energies to the band gap by dividing out $E_g$ then define a dimensionless matrix element
\begin{equation}
    M = \frac{\hbar}{m_0P}\hat{\mathbf{e}}\cdot\mathbf{p}
\end{equation}
so that 
\begin{equation}
    H_{mn} = \frac{eA_0}{2\hbar}PM.
\end{equation}
Plugging into the transition rate per unit volume then gives
    \begin{equation}
        W = \frac{2\pi}{\hbar^5}\left(\frac{e}{2}\right)^4\left|A_{01}A_{02}\right|^2\frac{1}{V}\sum_{\mathbf{k}_t}\sum_{cv}\left|\frac{M_{ci}^{2}M_{iv}^{1}}{E_{iv}-\hbar\omega_1}+\frac{M_{ci}^{1}M_{iv}^{2}}{E_{iv}-\hbar\omega_2}\right|^2 \delta(\hbar\omega_1 + \hbar\omega_2 - E_{cv}).
    \end{equation}
    At this stage the derivation is much simpler in atomic units where $\hbar = m_0 = e = 1/(4\pi\epsilon_0) = 1$. Converting units and normalizing all energies to the bandgap $E_g$ gives
    \begin{equation}
        W = 2\pi\frac{A_{01}^2A_{02}^2}{16}\frac{1}{V}\frac{P^4}{E_g^3}\sum_{\mathbf{k}_t}\sum_{cv}\left|\frac{M_{ci}^{2}M_{iv}^{1}}{\epsilon_{iv}-x_1}+\frac{M_{ci}^{1}M_{iv}^{2}}{\epsilon_{iv}-x_2}\right|^2 \delta(x_1 + x_2 - \epsilon_{cv}).
    \end{equation}
    We now convert the sum to an integral over the first Brillouin zone by
    \begin{equation}
        \frac{2\pi}{V}\sum_{\mathbf{k}_t} \rightarrow \frac{2\pi}{V}\int_{BZ}\frac{L_xL_y}{(2\pi)^2}d\mathbf{k}_t.
    \end{equation}
    Since the volume of the crystal $V_c = L_xL_yL_z$, we are left with (after changing to polar coordinates)
    \begin{equation}
        \frac{2\pi}{V_c}\sum_{\mathbf{k}_t} \rightarrow \frac{1}{L_z}\frac{1}{2\pi}\int_{0}^{2\pi}d\phi\int_{0}^{z.b.} k dk.
    \end{equation}
    Noting that $kP$ has units of energy, we can define another normalized parameter $\kappa = kP/E_g$. Changing the integral to $\kappa$ now gives
\begin{equation}
    W = \frac{A_{01}^2A_{02}^2}{16}\frac{P^2}{E_g}f_2\left(\frac{\hbar\omega_1}{E_g}; \frac{\hbar\omega_2}{E_g}\right),
    \end{equation}
    where
\begin{equation}
    f_2 = \sum_{cv}\frac{1}{2\pi}\int_{0}^{2\pi}d\phi\int_0^{z.b.}\kappa \left|\sum_i\frac{M_{ci}^{2}M_{iv}^{1}}{\epsilon_{iv}-x_1}+\frac{M_{ci}^{1}M_{iv}^{2}}{\epsilon_{iv}-x_2}\right|^2 \delta(x_1 + x_2 - \epsilon_{cv})d\kappa.
\end{equation}
Now we can convert to a 2PA coefficient by
\begin{equation}
    \alpha_2(x_1;x_2) = \frac{x_1 E_g}{2 I_1 I_2}W
\end{equation}
where the irradiance $I_l$ is given in atomic units as 
\begin{equation}
    I_l = \frac{c}{16\pi}n_l A_{0l}^2 x_l^2 E_g^2.
\end{equation}
Plugging this in and simplifying gives
\begin{equation}
    \alpha_2(\omega_1;\omega_2) = K\frac{E_p}{n_1 n_2 E_g^4 L_z}f_2\left(\frac{\hbar\omega_1}{E_g}; \frac{\hbar\omega_2}{E_g}\right),
\end{equation}
where $K$ is the material-independent parameter
\begin{equation}
    K = \left(\frac{\pi}{c}\right)^2 = \left(\frac{\pi}{137}\right)^2.
\end{equation}
Now, we can perform the delta function integration on the dimensionless spectral function using
\begin{equation}
    \int f(x)\delta[g(x)]dx = \sum_{x_0}f(x_0)\left|\frac{\partial{g}}{\partial{x}}\right|_{x_0}^{-1},
\end{equation}
where $x_0$ is a real solution to the equation $g(x) = 0$. Finally, we find that
\begin{equation}
    f_2 = \sum_{cv}\frac{1}{2\pi}\int_{0}^{2\pi}d\phi\sum_{\kappa_0}\kappa_0 \left|\frac{\partial{\epsilon_{cv}}}{\partial{\kappa}}\right|_{\kappa_0}^{-1}\left|\sum_i\frac{M_{ci}^{2}M_{iv}^{1}}{\epsilon_{iv}-x_1}+\frac{M_{ci}^{1}M_{iv}^{2}}{\epsilon_{iv}-x_2}\right|^2.
    \end{equation}
    
\section{Envelope functions and effective masses}
Here we show the origin of the energy dependent effective masses and envelope Schr{\"o}dinger equation. This derivation is not very different than the one in Ref.\,\cite{Bastard1976}, but we include it because it is expanded upon for the 14-band model. We begin with the expansion in zone center wave functions
\begin{equation}\label{eq:EnvelopeExp}
    \psi_{j n}(\mathbf{r}; \mathbf{k}_t) = e^{i\mathbf{k}_t\cdot \mathbf{r}}\sum_{\nu }\chi_{j n}^{\nu }(z; \mathbf{k}_t)u_{\nu  0}(\mathbf{r}),
\end{equation}
where $\chi^{\nu}_{jn}$ is the envelope component corresponding to the basis function $u_{\nu 0}(\mathbf{r})$. We will first use the 8 band basis that includes only two spin-degenerate conduction bands. Plugging Eq.\,(\ref{eq:EnvelopeExp}) into the time-independent Schr{\"o}dinger equation gives a new $\mathbf{k}\cdot\mathbf{p}$ form. Defining a vector $\boldsymbol{\chi}$ of the $8$ envelope functions, the result is found to be 
\begin{equation}
    \mathscr{D}\boldsymbol{\chi}  = E\boldsymbol{\chi},
\end{equation}
where $\mathscr{D}$ is the matrix \cite{Bastard1976}
\begin{align}\label{eq:D}
    \mathscr{D}_{\nu\nu'} &= \left(E_{n'} + \frac{\hbar^2k_t^2}{2m_0} + \frac{p_z^2}{2m_0} + V(z)\right)\delta_{\nu\nu'} \nonumber \\
                      &+  \frac{1}{m_0}p_{\nu\nu'}^zp_z + \frac{1}{2}p_z\frac{1}{\mathscr{M}_{\nu\nu'}^{zz}}p_z \nonumber \\
                      &+ \frac{\hbar}{m_0}\mathbf{k}_t\cdot \mathbf{p}_{\nu\nu'} + \frac{\hbar^2}{2}\sum_{\alpha,\beta = x,y}k_\alpha\frac{1}{\mathscr{M}_{\nu\nu'}^{\alpha \beta}}k_\beta \nonumber \\
                      &+ \frac{\hbar}{2}\sum_{\alpha = x,y}k_\alpha\frac{1}{\mathscr{M}_{\nu\nu'}^{\alpha z}}p_z + p_z \frac{1}{\mathscr{M}_{\nu\nu'}^{\alpha z}}k_\alpha.
\end{align}
The inverse effective mass terms $\mathscr{M}$ arise from perturbative inclusion of remote bands using L{\"o}wdin's method \cite{Lowdin1951, Chuang1995}. These perturbative coupling terms between bands $\nu$ and $\nu'$ are the tensors
\begin{equation}
    \frac{m_0}{\mathscr{M}_{\nu\nu'}^{\alpha\beta}} = \sum_{\lambda}\frac{\braket{\nu|p_\alpha|\lambda}\braket{\lambda|p_\beta|\nu'}}{E - E_\lambda - V_\lambda(z)}.
\end{equation}
We can choose $\mathbf{k}_t = 0$, which simplifies Eq.\,(\ref{eq:D}) considerably (the last two rows disappear). For $\alpha$ states we then have
\begin{align}
    \left(E_c + \frac{p_z^2}{2} + V_c(z) + \frac{1}{2}p_z \frac{1}{\mathscr{M}_{cc}^{zz}}p_z\right)\chi_c + p_{cl}^z p_z \chi_l + p_{cs}^z p_z \chi_s &= E\chi_c \nonumber \\
    \left(E_l + \frac{p_z^2}{2} + V_l(z) + \frac{1}{2}p_z \frac{1}{\mathscr{M}_{ll}^{zz}}p_z\right)\chi_l + p_{lc}^z p_z \chi_c + \frac{1}{2}p_z\frac{1}{\mathscr{M}_{ls}^{zz}}p_z \chi_s &= E\chi_l \nonumber \\
    \left(E_h + \frac{p_z^2}{2} + V_h(z) + \frac{1}{2}p_z \frac{1}{\mathscr{M}_{hh}^{zz}}p_z\right)\chi_h &= E\chi_h \nonumber \\
    \left(E_s + \frac{p_z^2}{2} + V_s(z) + \frac{1}{2}p_z \frac{1}{\mathscr{M}_{ss}^{zz}}p_z\right)\chi_s + p_{sc}^z p_z \chi_c + \frac{1}{2}p_z\frac{1}{\mathscr{M}_{sl}^{zz}}p_z \chi_l &= E\chi_s,
\end{align}
where the band index is now written with a subscript instead of a superscript. The equations for $\beta$ envelopes are identical. Note that the heavy hole envelope is decoupled from the rest of the basis, allowing the immediate reduction to a Schr{\"o}dinger equation for the single envelope
\begin{equation}\label{eq:Sh}
    \frac{1}{2}p_z\left(1 + \frac{1}{\mathscr{M}_{hh}^{zz}}\right)\chi_h + V_h(z)\chi_h = (E - E_h)\chi_h.
\end{equation}

We will assume all valence bands have the same offset $V_l = V_h = V_s = V_v$. The rest of the equations involve intra-basis coupling terms. We rearrange these remaining envelope equations and plug in the appropriate $p_{ij}^z$ to find
\begin{align}\label{eq:envelopes}
    \left[\frac{1}{2}p_z\left(1 + \frac{1}{\mathscr{M}_{ll}^{cc}}\right)p_z + V_c(z) \right]\chi_c + \sqrt{\frac{2}{3}} P p_z \chi_l - \sqrt{\frac{1}{3}} P p_z\chi_s &= (E - E_g)\chi_c \nonumber \\
    \left[\frac{1}{2}p_z\left(1 + \frac{1}{\mathscr{M}_{ll}^{zz}}\right)p_z + V_v(z) \right]\chi_l + \sqrt{\frac{2}{3}} P p_z \chi_c + \frac{1}{2}p_z\frac{1}{\mathscr{M}_{ls}^{zz}}p_z \chi_s &= E\chi_l \nonumber \\
    \left[\frac{1}{2}p_z\left(1 + \frac{1}{\mathscr{M}_{ss}^{zz}}\right)p_z + V_v(z) \right]\chi_s - \sqrt{\frac{1}{3}} P p_z \chi_c + \frac{1}{2}p_z\frac{1}{\mathscr{M}_{sl}^{zz}}p_z \chi_l &= (E + \Delta)\chi_s.
\end{align}
With the goal of reducing these to single envelope Schr{\"o}dinger equations as in Eq.\,(\ref{eq:Sh}), we would like to eliminate all other envelopes in favor of the dominant one. Clearly, the dominant envelope will be the one corresponding to the band's character. We assume slowly varying envelopes so we can neglect terms of order $p_z^2$ in comparison to first derivatives. Solving for the dominant envelope in each equation (in order) gives
\begin{align}\label{eq:subs}
    \chi_c &= \sqrt{\frac{2}{3}}P\frac{1}{E - E_g - V_c(z)}p_z\chi_l -  \sqrt{\frac{1}{3}}P\frac{1}{E - E_g - V_c(z)}p_z\chi_s \nonumber \\
    \chi_l &= \sqrt{\frac{2}{3}}P\frac{1}{E - V_v(z)}p_z\chi_c \nonumber \\
    \chi_s &= -\sqrt{\frac{1}{3}}P\frac{1}{E + \Delta - V_v(z)}p_z\chi_c.
\end{align}
Plugging in the expressions for $\chi_l$ and $\chi_s$ into the $\chi_c$ envelope equation (Eq.\,(\ref{eq:envelopes})) yields
\begin{align}
\frac{1}{2}p_z\left(1 + \frac{1}{\mathscr{M}_{cc}^{zz}} \right.&+\left. \frac{2}{3}\frac{E_p}{E - V_v(z)}  + \frac{1}{3}\frac{E_p}{E + \Delta - V_v(z)}\right)p_z \nonumber \\
    & + V_c(z) \chi_c = (E - E_g)\chi_c,
\end{align}
which is once again a single envelope equation. We eliminated $P$ using the fact that $E_p = 2P^2$ in atomic units. The effective mass term now has an energy-dependent portion due to coupling within the basis and an energy-independent portion term due to remote bands. Repeating for light hole envelopes, we find
\begin{align}
    &\left[\frac{1}{2}p_z\left(1 + \frac{1}{\mathscr{M}_{ll}^{zz}} + \frac{2}{3}\frac{E_p}{E - E_g - V_c(z)}\right)p_z + V_v(z) \right]\chi_l \nonumber \\
    & + \frac{1}{2}p_z\left(\frac{1}{\mathscr{M}_{ls}^{zz}} + \frac{\sqrt{2}}{3} \frac{E_p}{E - E_g - V_c(z)}\right)p_z \chi_s = E \chi_l
\end{align}
At first glance, this is more complicated than conduction band solution because it has not reduced to a single envelope equation. However, we can once again plug in the expression for $\chi_s$ as a function of $\chi_c$. Finally, we substitute for $\chi_c$ in the same way yielding terms of minimum order $p_z^4$. As always we ignore these relative to the second order. The conclusion is that we may ignore the effect of the $\chi_s$ envelope so that
\begin{equation}
    \left[\frac{1}{2}p_z\left(1 + \frac{1}{\mathscr{M}_{ll}^{zz}} + \frac{2}{3}\frac{E_p}{E - E_g - V_c(z)}\right)p_z + V_v(z) \right]\chi_l = E\chi_l.
\end{equation}
Repeating the exact same process for $\chi_s$ gives

\begin{equation}
    \left[\frac{1}{2}p_z\left(1 + \frac{1}{\mathscr{M}_{ss}^{zz}} + \frac{1}{3}\frac{E_p}{E - E_g - V_c(z)}\right)p_z + V_v(z) \right]\chi_s = E\chi_s.
\end{equation}

\subsection{14-band model}
The 14 band model comes from a relatively trivial extension of the 8 band analysis. The six added conduction bands have the same symmetry as the valence bands, so they are completely decoupled from each other to first order. Consequently, the valence bands derived in the 8 band model are equally valid in the 14 band case (within the approximations applied). The higher conduction bands have the expansion
\begin{align}\label{eq:}
    \left(E_{l'} + \frac{p_z^2}{2} + V_{l'}(z) + \frac{1}{2}p_z \frac{1}{\mathscr{M}_{l'l'}^{zz}}p_z\right)\chi_{l'} + p_{l'c}^z p_z \chi_c + \frac{1}{2}p_z\frac{1}{\mathscr{M}_{l's'}^{zz}}p_z \chi_{s'} &= E\chi_{l'} \nonumber \\
    \left(E_{s'} + \frac{p_z^2}{2} + V_{s'}(z) + \frac{1}{2}p_z \frac{1}{\mathscr{M}_{s's'}^{zz}}p_z\right)\chi_{s'} + p_{s'c}^z p_z \chi_c + \frac{1}{2}p_z\frac{1}{\mathscr{M}_{s'l'}^{zz}}p_z \chi_{l'} &= E\chi_{s'}.
\end{align}
Since we examine 2PA well below the energy of the higher conduction bands, we do not actually need effective mass relations for them. We just need to know their effect on the conduction band effective mass. To that end, we re-arrange as before to find
\begin{align}\label{eq:14exp}
    \chi_{l'} &= \sqrt{\frac{2}{3}}P\frac{1}{E - E_{l'} - V_{l'}(z)}p_z\chi_c \nonumber \\
    \chi_{s'} &= -\sqrt{\frac{1}{3}}P\frac{1}{E - E_{s'} - V_{s'}(z)}p_z\chi_c .
\end{align}
Finally, we look at the $\mathbf{k}\cdot\mathbf{p}$ expansion for the conduction band:
\begin{align}
    \left(E_c + \frac{p_z^2}{2} + V_c(z) + \frac{1}{2}p_z \frac{1}{\mathscr{M}_{cc}^{zz}}p_z\right)\chi_c &+ p_{cl}^z p_z \chi_l + p_{cs}^z p_z \chi_s  \nonumber \\ 
    &+ p_{cs'}^z p_z \chi_{s'} + p_{cl'}^z p_z \chi_{l'} = E\chi_c. \nonumber \\
\end{align}
Plugging in the expansions in Eq.\,(\ref{eq:14exp}) yields
\begin{equation}\label{eq:SEM}
    \frac{1}{2}p_z \frac{1}{m_{c}^z(E_{c},z)}p_z\chi_{c}(z) + V_{c}(z)\chi_{c}(z) = E_{c}\chi_{c}(z),
\end{equation}
with the effective mass altered to 
\begin{align}\label{eq:mc2}
    \frac{m_0}{m_c^z(E_c,z)} = 1 &+ \frac{m_0}{\mathscr{M}_{cc}^{zz}} \nonumber \\
    &+\frac{2}{3}\frac{E_p}{E - V_{l}(z)} + \frac{1}{3}\frac{E_p}{E + \Delta - V_s(z)} \nonumber \\
    &+ \frac{2}{3}\frac{E_p'}{E - E_{g}' - V_{l'}(z)} + \frac{1}{3}\frac{E_p'}{E - E_{s'} - V_{s'}(z)}.
\end{align}
\section{Analytical expression for normalized transmission}
Since the pump photon energy is too low to experience 2PA and we are ignoring the effects of nonlinear refraction, we take the pump propagation to be linear with loss $\sigma_2$ 
\begin{equation}\label{eq:linprop}
    \left(\frac{\partial}{\partial{x}} + \beta_1^{(2)} \frac{\partial{}}{\partial{t}} \right)A_2(x,t) = -\frac{\sigma_2}{2}A_2(x,t).
\end{equation}
The propagation direction is chosen along the $x$ axis to be consistent with the article. We can solve Eq.\,(\ref{eq:linprop}) by transforming into the moving frame where $T = t - \beta_1^{(2)} x$ and $X = x$. After converting back to the original coordinates, we see that the solutions must be of the form
\begin{equation}
    A_2(x,t) = F\left(0, t - \beta_1^{(2)} x\right)\exp\left(-\frac{\sigma_2}{2}x\right)
\end{equation}
Letting $A_2$ be Gaussian and remembering that $|A_l|^2 = P_l$, 
\begin{equation}\label{eq:P2}
    P_2 = |A_2|^2 = P_{02}\exp\left[-\frac{(t - \beta_1^{(2)} x)^2}{\tau_2^2}\right]\exp\left(-\sigma_2 x\right).
\end{equation}
Now we can plug into the equation for the probe traveling in the presence of the pump
\begin{equation}
    \left(\frac{\partial{}}{\partial{x}} + \beta_1^{(1)}\frac{\partial{}}{\partial{t}}\right)A_1 = 2i\gamma_{12}|A_2|^2 A_1 - \frac{\sigma_1}{2}A_1
\end{equation}
Next we can define an $A_1'$ such that $A_1 = A_1'\exp(-\sigma_1 x/2)$. Transform again to coordinates moving with the probe group velocity $T = t - \beta_1^{(1)}x$ and $X = x$, giving 
\begin{equation}
    \frac{\partial{A_1'(X,T)}}{\partial{X}} = 2i \gamma_{12} |A_2(X,T)|^2 A_1'(X,T).
\end{equation}
We can calculate the power evolution of wave $1$ by
\begin{equation}\label{eq:PP}
    \frac{\partial{}}{\partial{X}}|A_1'|^2 = \frac{\partial{P_1}}{\partial{x}} = A_1^{'*}\frac{\partial{A_1'}}{\partial{X}} + A_1'\frac{\partial{A_1^{'*}}}{\partial{X}} = -4\mathrm{Im}\{\gamma_{12}\}P_2 P_1.
\end{equation}
We take the probe input to be Gaussian with a time delay of $\tau$ relative to the pump:
\begin{equation}
    P_1(X,T) = P_{01}(X)\exp\left[-\frac{(T - \tau)^2}{\tau_1^2}\right].
\end{equation}
Using the pump distribution from Eq.\,(\ref{eq:P2}) with the substitution $t = T + \beta_1^{(1)}$ and plugging into Eq.\,(\ref{eq:PP}) along with the probe profile, we find 
\begin{align}
    \frac{\partial{E_1(X)}}{\partial{X}} &= -4 \mathrm{Im}\{\gamma_{12}\} P_{01}(X) P_{02} \exp\left(-\sigma_2 X\right)\nonumber \\
                                      &\times\int_{-\infty}^{\infty} \exp\left[-\frac{(T + \rho X)^2}{\tau_2^2}\right]\exp\left[-\frac{(T - \tau)^2}{\tau_1^2}\right] dT \nonumber \\
                                      &= -4 \mathrm{Im}\{\gamma_{12}\} P_{01}(X) P_{02} \sqrt{\pi}\frac{\tau_1 \tau_2}{\tau_x}\exp\left[-\frac{(\tau + \rho X)^2}{\tau_x^2} - \sigma_2 X\right].
\end{align}
The symbol $\rho = \beta_1^{(1)} - \beta_1^{(2)}$ is the GVM parameter. We can simplify a bit by noting that $E_j = \sqrt{\pi}\tau_j P_{0j}$, then integrate over the length of the waveguide to find the output energy
\begin{equation}
    \int_0^{L} \frac{d E_1}{E_1} = -\frac{4}{\sqrt{\pi}}\mathrm{Im}\{\gamma_{12}\} E_2 \frac{1}{\tau_x}\int_{0}^L \exp\left[-\frac{(\tau + \rho X)^2}{\tau_x^2} - \sigma_2 X\right] dX.
\end{equation}
The result is that
\begin{align}\label{eq:analytic}
    \log \left(\frac{E_1(L)}{E_1(0)}\right) &=  2\frac{E_2 \mathrm{Im}\{\gamma\}}{\rho}\exp{\left[\frac{\sigma_2 \tau}{\rho} + \left(\frac{\sigma_2\tau_x}{2\rho}\right)^2\right]}  \nonumber \\
             &\times \left[\mathrm{erf}\left(\frac{\tau}{\tau_x} + \frac{\sigma_2 \tau_x}{2\rho}\right) - \mathrm{erf}\left(\frac{\tau + \rho L}{\tau_x} + \frac{\sigma_2 \tau_x}{2\rho}\right)\right]
\end{align}
Recall we are working with $A'$ which has the exponential decay factored out. However, the linear decay of the probe pulse does not affect the normalized transmission signal. It is defined as $E_1(L, E_2)/E_1(L, 0)$, or the ouput energy with the pump divided by the output energy without the pump. Since both will gain a factor of $\exp(-\sigma_1 L)$, we may ignore it.